\documentclass[a4paper,11pt]{article}

\usepackage{jheppub}
\usepackage[T1]{fontenc}
\usepackage{float}

\title{\textbf{Leptogenesis, Fermion
		Masses and Mixings in a SUSY $SU(5)$ GUT with $D_{4}$ Flavor
		Symmetry}}

\author[a]{M. Miskaoui}
\author[a]{and M. A. Loualidi}

\affiliation[a]{LPHE, Modeling and Simulations, Faculty of Science,
	Mohammed V University in Rabat, 10090 Rabat, Morocco}

\emailAdd{m.miskaoui@gmail.com}
\emailAdd{mr.medamin@gmail.com}

\abstract{We propose a model of fermion masses and mixings based on $SU(5)$
	grand unified theory (GUT) and a $D_{4}$ flavor symmetry. This is a highly predictive 4D $SU(5)$ GUT with a flavor
	symmetry that does not contain a triplet irreducible representation.
	The Yukawa matrices of quarks and charged leptons are obtained after
	integrating out heavy messenger fields from renormalizable
	superpotentials while neutrino masses are originated from the type I
	seesaw mechanism. The group theoretical factors from 24- and
	45-dimensional Higgs fields lead to ratios between the Yukawa
	couplings in agreement with data, while the dangerous proton decay
	operators are highly suppressed. By performing a numerical fit, we
	find that the model captures accurately the mixing angles, the
	Yukawa couplings and the $CP$ phase of the quark sector at the GUT
	scale. The neutrino masses are generated at the renormalizable level
	with the prediction of trimaximal mixing while an additional
	effective operator is required to account for the baryon asymmetry
	of the universe (BAU). The model is remarkably predictive because
	only the normal neutrino mass ordering and the lower octant of the
	atmospheric angle are allowed while the $CP$ conserving values of
	the Dirac neutrino phase $\delta_{CP}$ are excluded. Moreover, the
	predicted values of the effective Majorana mass $m_{\beta \beta }$
	can be tested at future neutrinoless double beta decay experiments.
	An analytical and a numerical study of the BAU via the leptogenesis
	mechanism is performed. We focused on the regions of parameter space
	where leptogenesis from the lightest right-handed neutrino is
	successfully realized. Strong correlations between the parameters of
	the neutrino sector and the observed BAU are obtained.}


\begin{document}
\maketitle	
\flushbottom	

\section{Introduction}
During the last two decades, neutrino oscillation experiments have presented
vigorous measurements of the neutrino mass-squared differences and their
mixing angles\footnote{%
	The global analysis of all available oscillation data can be found in \cite%
	{R3,R4,R5}} \cite{R1,R2}, conflicting the zero mass prediction of
the standard model (SM) of electroweak interactions. Besides the
precise measurement of the oscillation parameters, the $CP$
violation and the flavor pattern in the quark and lepton sectors
along with the fermion mass hierarchies are not firmly established
within the SM. Therefore, theoretical investigations beyond the SM
are urgently needed to explain the fermion flavor structure. The $CP$ violation is of particular interest especially after the
T2K collaboration excluded some values of $\delta _{CP}$ giving rise
to a large improvement of the observed antineutrino oscillation
probability at $3\sigma $ confidence level \cite{R6}. Moreover, $CP$
violation is one of the essential ingredients among the three
conditions presented by Sakharov to explain the observed BAU through
the baryogenesis mechanism \cite{R7}. The other two conditions being
the baryon number violation and the deviation from thermal
equilibrium. The reason to search for this in the lepton sector is
due to the fact that the SM predictions for the $CP$ violation --
which is encoded in the CKM phase $\delta _\mathrm{CKM}$ -- is
insufficient to generate the observed BAU and thus, new sources for
$CP$ violation beyond the SM are required. An interesting approach to successfully produce the observed excess of
matter over antimatter in the universe is through the leptogenesis mechanism\footnote{%
	There are several baryogenesis models using different scenarios to explain
	the BAU, for a review see refs. \cite{R8,R9,R10,R11,R12}.}
\cite{R13}, which relies on the right-handed (RH) Majorana neutrinos
introduced in the context of type I seesaw mechanism \cite%
{R14,R15,R16,R17,R18}. In practical terms, this approach requires
lepton number violation which arise naturally in type I seesaw
models via the Majorana masses of the RH neutrinos. Then, a lepton
asymmetry is generated by the out of-thermal-equilibrium and $CP$ violating
decays of these RH neutrinos that is eventually converted into a
primordial baryon asymmetry by means of the SM sphaleron processes
\cite{R19}. As a result, the three Sakharov conditions are satisfied
in this scenario, which is remarkable considering that leptogenesis
connects high energy scales where the BAU
takes place and neutrino oscillations that take place at low energy scales.%
\newline
\newline
Grand unified theories are the most attractive high-energy
completions of the SM that can bridge the experimentally accessible
low energies with extremely high energy phenomenon while providing the unification of electromagnetic, weak and strong interactions
\cite{R20,R21,R22,R23,R24}. When combined with
supersymmetry (SUSY) \cite{R25}, GUTs provide a more powerful
explanation to some of the open questions in the SM as well as a
solution to some of the problems that are not addressed in the
minimal non-SUSY GUTs \cite{R26}. The simplest realization of such a
combination is provided by the SUSY $SU(5)$ model where the unification of the gauge couplings occurs at a scale
of approximately $2\times 10^{16}$ $\mathrm{GeV}$ \cite{R27,R28,R29,R30,R31}. As a result of unification, the masses of down quarks and charged leptons
are generated from a common renormalizable operator leading to Yukawa couplings of same order of magnitude; $%
y_{e}=y_{d},$ $y_{\mu }=y_{s},$ and $y_{\tau }=y_{b}$. It is
well-known that these equalities are acceptable for the third
generation but fails for the remaining ones because of their
conflict with the experimental data. The Georgi Jarlskog (GJ)
relations $m_{\mu }/m_{s}=3$ and $m_{e}/m_{d}=1/3$ generated from a specific renormalizable operator involving
a 45-dimensional Higgs $H_{45}$ presented a first example solution
to this issue \cite{R32}. In contrast to these relations, considering
additional Higgs fields in the 24- or 75-dimensional
representations of $SU(5)$ gives rise to nontrivial Clebsch-Gordan (CG) factors with
new ratios for the first two generations of Yukawa couplings
that are preferred phenomenologically; see for instance refs.
\cite{R39,R40} for ratios derived from dimension 5 and dimension 6
operators in the context of SUSY $SU(5)$. On the other hand, neutrinos are massless in SUSY $SU(5)$ model
which implies that the oscillation phenomenon can not be explained
within its minimal realization. The simplest way to address this
issue is by introducing RH singlet fermions to generate neutrino
masses via the type I seesaw mechanism, while the mixing angles can
be determined by invoking the
well-known approach of flavor symmetries. Non-Abelian discrete symmetries%
\footnote{%
	For reviews on the use of non-Abelian discrete symmetries, see for instance
	\cite{A1,A2,A3,A4,A5}.} are in particular a powerful tool for explaining the mass hierarchies and the mixing of all
fermions \cite{A6}, especially, those with triplet representations. For example, the discrete symmetry $%
A_{4}$ is widely used in $SU(5)$ flavor models to explain the
patterns of
neutrino masses and their mixing; see for instance refs. \cite%
{A6,B1,B2,B3,B4,B7,B8,B10,B11}. On the other hand, the discrete groups
with doublet representations like $S_{3}$ and $D_{4}$ are less
employed in 4D $SU(5)$ GUTs. In fact, a SUSY $SU(5)$ model based on $D_{4}$ symmetry was considered before in ref. \cite{D1}; however, the phenomenological implications of both the lepton and quark sectors were lacking. Here, we will show that $D_{4}$ can provide good results regarding
the neutrinos as well as the charged fermions flavor structures by allowing the three generations of matter to be unified into the
representations $1$ and $2$ instead of $3$\footnote{%
	As shown in Ref. \cite{B12}, one of the interesting properties of
	models with the $D_{4}$ group is that it predicts the well-known $\mu -\tau $
	symmetry in a natural manner \cite{B13,B14,B15,B16,B17} by using
	minimal set of flavon fields.}.\newline
\newline
In this work, we build a predictive model based on SUSY $SU(5)$ GUT
supplemented by a $D_{4}$ flavor symmetry suitable for addressing
the above mentioned questions. In particular, we show that our
construction leads to results for the pattern of fermion masses and
mixings that are consistent with the current experimental data. In
fact, this is the first phenomenological analysis of the fermion
mass and mixing structures within
SUSY $SU(5)$ using the dihedral group $D_{4}$. Besides this discrete group, we have added a $U(1)$
symmetry to engineer the invariance of the superpotentials in the
quark and lepton sectors, and also to prevent dangerous operators
that mediate rapid proton decay. Apart from the usual SUSY $SU(5)$
superfield spectrum, various superfields are added to the model in order to fulfill different tasks. Namely, many messenger fields denoted as $X_{i}$ and $Y_{i}$ are needed to make the model renormalizable, higher dimensional Higgs fields in the 24 and 45 representations required to obtain realistic Yukawa coupling ratios, gauge singlets superfields -- the so-called flavons -- needed to break the flavor symmetry and structure the fermions mass
matrices, and three right-handed neutrinos
$N_{i=1,2,3}^{c}$ responsible for the tiny neutrino
masses as well as the BAU through
the leptogenesis mechanism\footnote{For leptogenesis models based on
	type I seesaw mechanism in the
	framework of $SU(5)$ GUT see, for instance, ref. \cite%
	{A8,A8b,A9,A10,A11} and references therein.}. The introduction of all
of the above fields with the requirement to keep the effective
superpotentials
invariant is highly controlled by the group theoretical structure of the $%
D_{4}\times U(1)$ flavor symmetry.\newline
In the charged sector, the messenger fields $X_{i}$ and $Y_{i}$ are
coupled to the matter fields, the flavon fields and the 24 and 45
Higgs fields. When $X_{i}$ and $Y_{i}$ are integrated out we obtain
the effective operators responsible for the quark and lepton Yukawa
couplings, which is then followed by the spontaneous breaking of the
flavor and gauge
symmetries after the flavons and Higgs fields of the $SU(5)$ -- that is the $%
5 $, $\bar{5}$, $24$, and $45$\ dimensional Higgs fields denoted
respectively as $H_{5}$, $H_{\bar{5}}$, $H_{24}$ and
$H_{45}$ -- acquire nonzero vacuum expectation values (VEVs). On the
one hand, the specific VEV alignments of the flavons break the
$D_{4}\times U(1)$ symmetry and help shape the fermions mass
matrices, leading eventually to the appropriate flavor structure of
the quarks and leptons. On the other hand, the CG factors obtained
from the VEV structures of $H_{24}$ and $H_{45}$ lead to the
following double ratio of the Yukawa coupling of the first and
second generation $\frac{y_{\mu }}{y_{s}}\frac{y_{d}}{y_{e}}\simeq
10.12$ which is consistent with experimental data
\cite{A12}.
\newline
In the chargeless sector, the neutrino masses are generated at the
renormalizable level through the type I seesaw mechanism. The obtained neutrino mass matrix $%
m_{\nu }$ is described by only three parameters leading to strong
constraints among the physical parameters. Moreover, $m_{\nu }$ is
invariant under a particular remnant $Z_{2}$ symmetry which is
commonly referred to as a magic symmetry \cite{A13}, indicating that
$m_{\nu }$ is diagonalized by the well-known trimaximal mixing
($TM_{2}$) matrix which is consistent with the observed neutrino mixing angles \cite%
{A14,A15,A16,A17,A18,A19,A20}. However, the leptogenesis mechanism can not
be induced at the renormalizable level given that the neutrino Yukawa
coupling matrix is proportional to the identity matrix which leads to a
vanishing lepton asymmetry. Therefore, we show that by introducing one
effective operator as a correction to the neutrino Yukawa coupling
matrix, our model can accommodate successfully the observed BAU via
leptogenesis\footnote{%
	This is a known requirement in models predicting the tribimaximal
	mixing (TBM) at the leading order, see for instance refs. \cite{A21,A22}.}. Our main results in the neutrino
sector are:

\begin{itemize}
	\item only the normal hierarchy (NH) for neutrino mass spectrum is allowed,
	
	\item only the lower octant of the atmospheric angle is allowed,
	
	\item the $CP$ conserving values of the Dirac $CP$ phase $\delta _{CP}$ are
	excluded,
	
	\item the predicted values of the effective Majorana mass in neutrinoless
	double beta decay ($0\nu \beta \beta $) are testable at future $0\nu \beta
	\beta $ searches, and
	
	\item the correlation between the BAU parameter denoted as $Y_{B}$ and
	the neutrino sector parameters satisfies the experimental bound of
	the baryon asymmetry from the Planck collaboration \cite{A23}.
\end{itemize}

The rest of the paper is organized as follows. In section
\ref{sec2}, we present the particle content of the model as well as
their transformation
properties under the $SU(5)\times D_{4}\times U(1)$ symmetry. In section \ref%
{sec3}, we derive the mass matrices of the charged fermions and give
brief comments on the fast proton decay operators within our
construction. In section \ref{sec4}, we study the neutrino sector
where the analytical expressions of the neutrino masses and mixing
parameters are obtained as a function of the model parameters. In
section \ref{sec5}, we show that a perfect fit to the
fermion masses and mixings can be obtained for all observables. In section %
\ref{sec6}, we carry out an analytical and a numerical study of the
BAU via the leptogenesis mechanism. A conclusion is given in section
\ref{sec6}. Appendix \ref{app1} describes the messenger sector of
the model. Appendix \ref{app2} shows that the contribution of the
charged leptons and the higher dimensional Dirac operators to the lepton asymmetry is highly suppressed to
account for the BAU. Appendix \ref{app3} provides some tools on
$D_{4}$ discrete group. Appendix \ref{app4} describes the realization of the vacuum alignment of $D_4$ flavon doublets. 
\section{Theoretical setup}
\label{sec2}
In this section, we describe the different sectors of our $SU(5)\times
D_{4}\times U(1)$ GUT proposal and fix some notations. The chiral sector of
the minimal supersymmetric $SU(5)$ model involves matter and Higgs
superfields which are both supplemented by extra superfields in the present
setup. Apart from the usual SUSY $SU(5)$
superfield spectrum, the building blocks of the
present model can be classified into four sets:
\begin{itemize}
	\item \emph{(a)} a
	renormalizable messenger sector with messenger fields $X_{i}$
	associated to down quarks, charged leptons and neutrinos, and
	$Y_{i}$ associated to the up quarks; details on this sector is
	provided in appendix \ref{app1},
	\item \emph{(b)} two
	additional higher dimensional Higgs fields in the 24 and 45 GUT
	representations required for gauge symmetry breaking and for generating Yukawa coupling ratios compatible with the data,
	\item \emph{(c)}
	several flavon superfields carrying
	quantum numbers under the flavor symmetry $D_{4}\times U(1) $
	needed to break the flavor symmetry and structure the fermions mass
	matrices, and
	\item \emph{(d)} three right-handed neutrinos
	$N_{i=1,2,3}^{c}$ responsible for generating the tiny neutrino
	masses via the type I seesaw mechanism as well as the BAU through
	the leptogenesis mechanism.
\end{itemize} Recall that the usual matter superfields denoted as $%
T_{i}=(u_{i}^{c},e_{i}^{c},Q_{iL})$ and $F_{i}=(d_{i}^{c},L_{i})$ -- with $%
i=1,2,3$ refers to the three generations of matter -- fit into the $10_{i}$
and $\bar{5}_{i}$ representations respectively. Recall also that the low
energy Higgs doublets $H_{u}$ and $H_{d}$ of the minimal
supersymmetric standard model (MSSM) arise from $H_{5}$ and a
mixture of $H_{\overline{5}}$ and $H_{\overline{45}}$ respectively.
The 45-dimensional Higgs is usually used to produce the GJ relations
differentiating between the (2-2) entry of the down quark and
charged lepton mass matrices; however, it has been shown
in \cite{R39,R40} that there are many other options which are preferred compared to GJ relations. These alternatives arise from higher-dimensional operators involving essentially
higher dimensional Higgs representations. In our proposal, we use $H_{%
	\overline{45}}$ and $H_{24}$ to produce the following ratios of the diagonal
Yukawa couplings $y_{e}/y_{d}=4/9$ and $y_{\mu }/y_{s}=9/2$ which are in perfect
agreement with experimental data \cite{A12}. The different steps leading to these
ratios is elaborated in the next section. The 45-dimensional Higgs
$H_{\overline{45}}$ satisfy the following relations%
\begin{eqnarray}
	(H_{\overline{45}})_{c}^{ab} &=&-(H_{\overline{45}})_{c}^{ba}\quad ,\quad
	(H_{\overline{45}})_{a}^{ab}=0  \notag \\
	\left\langle (H_{\overline{45}})_{i}^{i5}\right\rangle &=&\upsilon
	_{45}\quad ,\quad \left\langle (H_{\overline{45}})_{4}^{45}\right\rangle
	=-3\upsilon _{45}\ \text{with }i=1,2,3  \label{h45}
\end{eqnarray}%
where $\upsilon _{45}$ is the VEV of $H_{\overline{45}}$. As for the adjoint
Higgs $H_{24}$ which is also responsible for breaking the $SU(5)$ group, it
develops its VEV along the direction
\begin{equation}
	\left\langle (H_{24})_{b}^{a}\right\rangle =\mathrm{diag}(1,1,1,\frac{-3}{2},%
	\frac{-3}{2})\upsilon _{24}  \label{h24}
\end{equation}%
All the above superfields carry as well quantum numbers under the $%
D_{4}\times U(1)$ group as depicted in table (\ref{t1}). In this table, $%
F_{2,3}$ and $N_{3,2}^{c}$ notations stand for $D_{4}$ doublet assignments $%
(F_{2},F_{3})^{T}$\ and $(N_{3},N_{2})^{T}$ respectively.
\begin{table}[h]
	\centering
	\begin{tabular}{l||l|l|l|l|l|l|l|l|l|l|l}
		\hline
		& $T_{1}$ & $T_{2}$ & $T_{3}$ & $F_{1}$ & $F_{2,3}$ & $N_{1}^{c}$ & $%
		N_{3,2}^{c}$ & $H_{5}$ & $H_{\overline{5}}$ & $H_{\overline{45}}$ & $H_{24}$
		\\ \hline
		$SU(5)$ & $10_{1}$ & $10_{2}$ & $10_{3}$ & $\overline{5}_{1}$ & $\overline{5}%
		_{2,3}$ & $1_{1}^{\nu }$ & $1_{3,2}^{\nu }$ & $5_{H_{u}}$ & $\overline{5}%
		_{H_{d}}$ & $\overline{45}_{H}$ & $24_{H}$ \\ \hline
		$D_{4}$ & $1_{+,-}$ & $1_{+,-}$ & $1_{+,+}$ & $1_{+,+}$ & $2_{0,0}$ & $%
		1_{+,+}$ & $2_{0,0}$ & $1_{+,+}$ & $1_{+,-}$ & $1_{+,+}$ & $1_{+,+}$ \\
		\hline
		$U\mathbb{(}1\mathbb{)}$ & $6$ & $12$ & $4$ & $13$ & $13$ & $-5$ & $-5$ & $%
		-8 $ & $4$ & $-16$ & $0$ \\ \hline
	\end{tabular}%
	\caption{The $SU(5)\times D_{4}$ representations and $U(1)$ charges of the
		matter, RH neutrinos and Higgs superfields.}
	\label{t1}
\end{table}
On the other hand, the $D_{4}\times U(1)$ invariance requires the
introduction of several flavon fields in all the sectors of the
model. In the up-quark sector, only the top quark mass arises from a
tree level Yukawa coupling, the up and charm quark masses are
derived from higher dimensional couplings involving five flavon
fields denoted as $\xi _{i=1,...,5}$.
In the down quark and charged lepton sector, four flavon fields denoted as $%
\phi $, $\varphi $, $\Omega $ and $\Phi $ are needed for $%
D_{4}\times U(1)$ invariance. When these flavons acquire their
VEVs, they break the $D_{4}$ group and lead to appropriate mass matrices of
down quarks and charged leptons.
\begin{table}
	\centering
	\begin{tabular}{l||l|l|l|l|l||l|l|l|l}
		\hline
		Flavons & $\xi _{1}$ & $\xi _{2}$ & $\xi _{3}$ & $\xi _{4}$ & $\xi _{5}$ & $%
		\phi $ & $\varphi $ & $\Omega $ & $\Phi $ \\ \hline
		$D_{4}$ & $1_{+,+}$ & $1_{+,+}$ & $1_{+,-}$ & $1_{+,+}$ & $1_{+,-}$ & $%
		1_{+,-}$ & $1_{+,-}$ & $2_{0,0}$ & $2_{0,0}$ \\ \hline
		$U\mathbb{(}1\mathbb{)}$ & $-4$ & $-10$ & $-2$ & $-16$ & $-8$ & $-9$ & $-14$
		& $-21$ & $-9$ \\ \hline
	\end{tabular}%
	\caption{The $D_{4}\times U(1)$ quantum numbers of the flavons used in the
		quark and charged lepton sectors.}
	\label{t2}
\end{table}
In the neutrino sector, five flavons are required for $U(1)$
invariance. Three of them, denoted as $\rho _{1}$, $\rho _{2}$ and $%
\rho _{3}$, are assigned into different $D_{4}$\ singlets while the
remaining two denoted as $\digamma $ and $\Gamma $ are transforming as $%
D_{4} $ doublets.
\begin{table}[h]
	\centering
	\begin{tabular}{l||l|l|l|l|l}
		\hline
		Flavons & $\rho _{1}$ & $\rho _{2}$ & $\rho _{3}$ & $\digamma $ & $\Gamma $
		\\ \hline
		$D_{4}$ & $1_{+,+}$ & $1_{+,-}$ & $1_{-,-}$ & $2_{0,0}$ & $2_{0,0}$ \\ \hline
		$U\mathbb{(}1\mathbb{)}$ & $10$ & $10$ & $10$ & $10$ & $10$ \\ \hline
	\end{tabular}%
	\caption{The $D_{4}\times U(1)$ quantum numbers of the flavons used in the
		neutrino sector.}
	\label{t3}
\end{table}%
The flavons $\rho _{1}$ and $\digamma $ lead to the popular
tribimaximal mixing matrix \cite{C0}, while $\rho _{2}$, $\rho
_{3}$ and $\Gamma $ are responsible for the deviation of the
neutrino mixing angles from their TBM values. The quantum numbers
under $D_{4}\times U(1)$ of these five flavons is as depicted in
table (\ref{t3}).
\section{Charged fermion sector}
\label{sec3}
To derive the Yukawa matrices of the charged fermion sector, we
start by the
up-type quarks Yukawa matrix which descend from the trilinear interaction terms $%
10_{i}.10_{j}.5_{H_{u}}\equiv T_{i}T_{j}H_{5}$ where $i,j=1,2,3$ .
However, the up-type quarks Yukawa matrix is generated within our
construction from higher order operators derived from several
renormalizable terms involving messenger fields $Y_{i}$ and gauge
singlet flavon fields $\xi _{i}$, see appendix \ref{app1} for more
details on $Y_{i}$ and tables (\ref{t1}) and (\ref{t2}) to check the
invariance under $SU(5)\times D_{4}\times U(1)$ symmetry. After
integrating out these messenger fields\footnote{%
	The renormalizable superpotentials for the quarks before integrating
	out the messenger fields are given in appendix \ref{app1}.} we
obtain the invariant
effective superpotential for the up quarks%
\begin{eqnarray}
	W_{up} &=&\frac{y_{11}^{u}}{\Lambda }T_{1}T_{1}H_{5}\xi _{1}+\frac{y_{12}^{u}%
	}{\Lambda }T_{1}T_{2}H_{5}\xi _{2}+\frac{y_{13}^{u}}{\Lambda }%
	T_{1}T_{3}H_{5}\xi _{3}+\frac{y_{22}^{u}}{\Lambda }T_{2}T_{2}H_{5}\xi _{4}
	\notag \\
	&&+\frac{y_{23}^{u}}{\Lambda }T_{2}T_{3}H_{5}\xi
	_{5}+y_{33}^{u}T_{3}T_{3}H_{5}  \label{wu}
\end{eqnarray}%
where $y_{ij}^{u}$ are the Yukawa coupling constants and $ \Lambda
$ is the cutoff scale of the model which we take as the GUT scale.
The $D_{4}$ flavor symmetry is broken by the VEVs of the flavon
fields as $\left\langle \xi _{i}\right\rangle =\upsilon _{\xi _{i}}$
with $i=1,...,5$ while the
electroweak doublet $H_{u}$ contained in $H_{5}$ acquire its VEV as usual $%
\left\langle H_{5}\right\rangle =\upsilon _{u}$. Assuming that the
parameters in $W_{up}$ are all real, the Yukawa matrix of up-type quarks can
be written as%
\begin{equation}
	\mathcal{Y}_{up}=\left(
	\begin{array}{ccc}
		\frac{y_{11}^{u}}{\Lambda }\upsilon _{\xi _{1}} & \frac{y_{12}^{u}}{\Lambda }%
		\upsilon _{\xi _{2}} & \frac{y_{13}^{u}}{\Lambda }\upsilon _{\xi _{3}} \\
		\frac{y_{12}^{u}}{\Lambda }\upsilon _{\xi _{2}} & \frac{y_{22}^{u}}{\Lambda }%
		\upsilon _{\xi _{4}} & \frac{y_{23}^{u}}{\Lambda }\upsilon _{\xi _{5}} \\
		\frac{y_{13}^{u}}{\Lambda }\upsilon _{\xi _{3}} & \frac{y_{23}^{u}}{\Lambda }%
		\upsilon _{\xi _{5}} & y_{33}^{u}%
	\end{array}%
	\right) =\left(
	\begin{array}{ccc}
		a_{11} & a_{12} & a_{13} \\
		a_{12} & a_{22} & a_{23} \\
		a_{13} & a_{23} & a_{33}%
	\end{array}%
	\right)  \label{yu}
\end{equation}

We will now proceed with the down-type quarks and charged leptons generated
from the same Yukawa coupling $10_{i}.\overline{5}_{j}.\overline{5}%
_{H_{d}}\equiv T_{i}F_{j}H_{\overline{5}}$ where $i,j=1,2,3$ are the
generation indices. The superpotential leading to the Yukawa matrices is
obtained from a renormalizable superpotential that contains messenger fields
denoted as $X_{i}$. However, this time we need to add higher-dimensional
Higgs representations to differentiate between down quarks and charged
lepton masses, in particular we use the adjoint Higgs $H_{24}$ and the
45-dimensional Higgs $H_{\overline{45}}$ for this purpose, see eqs. (\ref%
{h45}) and (\ref{h24}). Therefore, by using the superfield
assignments in tables (\ref{t1}) and (\ref{t2}) and integrate out
the messenger fields, we get the following invariant effective
superpotential for the down
quarks and charged leptons%
\begin{equation}
	W_{e,d}=\frac{y_{11}^{d}}{\left\langle H_{24}\right\rangle ^{2}}%
	T_{1}F_{1}\phi \varphi H_{\overline{5}}+\frac{y_{12}^{d}}{\left\langle
		H_{24}\right\rangle ^{2}}T_{1}F_{2,3}\Phi \varphi H_{\overline{5}}+\frac{%
		y_{22}^{d}}{\Lambda ^{2}}T_{2}F_{2,3}\Phi H_{24}H_{\overline{45}}+\frac{%
		y_{33}^{d}}{\Lambda }T_{3}F_{2,3}\Omega H_{\overline{5}}  \label{wd}
\end{equation}%
where $y_{ij}^{d}$ are the Yukawa coupling constants associated to the down
quarks and charged leptons. To illustrate how the adjoint Higgs contributes
to the entries of the Yukawa matrices and leading subsequently to a
particular CG factors that distinguish the down quarks Yukawa couplings from
those of the charged leptons,let us discuss the $%
y_{11}^{d}$ effective operator in (\ref{wd}). This term is achieved by
integrating out the heavy messenger fields from the following renormalizable
terms\footnote{%
	The coupling constants are omitted in $W_{e,d}^{Ren}$ for clarity.}%
\begin{equation}
	W_{e,d}^{Ren}\supset F_{1}\phi X_{1}+\overline{X}_{1}\varphi X_{2}+\overline{%
		X}_{2}H_{\overline{5}}T_{1}+X_{1}H_{24}\overline{X}_{1}+X_{2}H_{24}\overline{%
		X}_{2}  \label{rwd}
\end{equation}%
After integrating out $X_{1,2}$ and $\overline{X}_{1,2}$ from the first
three terms in (\ref{rwd}), we are left with the first operator in (\ref{wd}%
); $T_{1}F_{1}\phi \varphi H_{\overline{5}}$. On the other hand, the last
two terms in eq. (\ref{rwd}) are responsible for the appearance of the square of
the Higgs adjoint VEV $\left\langle H_{24}\right\rangle ^{2}$ in the
denominator of the effective operator. Specifically, the masses of the
messenger pairs $X_{1,2}$ and $\overline{X}_{1,2}$ are achieved when $H_{24}$
acquire its VEV\footnote{%
	We assume for simplicity that the VEV of the adjoint Higgs is around the GUT
	scale as well as the cutoff scale; $\Lambda \equiv \left\langle
	H_{24}\right\rangle =M_{GUT}\simeq 2\times 10^{16}\mathrm{GeV}$.} $%
\left\langle H_{24}\right\rangle $ with the group structure given in eq. (%
\ref{h24}) which is then followed by integrating out $X_{1}%
\overline{X}_{1}$ and $X_{2}\overline{X}_{2}$ to obtain eventually
the first effective operator in (\ref{wd}). According to the group
structure of $\left\langle H_{24}\right\rangle $, the down quark
mass is multiplied by the inverse of the CG factors in the first
three entries of the adjoint Higgs VEV in eq. (\ref{h24}) which is
just $1$ in this case, while the electron mass is multiplied by the
inverse of the fourth and fifth components such that the resulting
CG coefficient is $\frac{-2}{3}\times \frac{-2}{3}=\frac{4}{9}$. The
same discussion holds for the second effective operator in
(\ref{wd}) while for the third effective operator the CG factors
arise in the numerator. For completeness, when the flavon fields
acquire their VEVs in
accordance with the following alignment%
\begin{equation}
	\left\langle \phi \right\rangle =\upsilon _{\phi }\quad ,\quad \left\langle
	\varphi \right\rangle =\upsilon _{\varphi }\quad ,\quad \left\langle \Phi
	\right\rangle =(\upsilon _{\Phi },0)^{T}\quad ,\quad \left\langle \Omega
	\right\rangle =(0,\upsilon _{\Omega })^{T} \label{vq}
\end{equation}%
we end up with the Yukawa matrices of the down-type quarks
$\mathcal{Y}_{d}$ and charged leptons $\mathcal{Y}_{e}$ expressed
as
\begin{eqnarray}
	\mathcal{Y}_{d} &=&\left(
	\begin{array}{ccc}
		\frac{y_{11}^{d}\upsilon _{\varphi }\upsilon _{\phi }}{\upsilon _{24}^{2}} &
		\frac{y_{12}^{d}\upsilon _{\varphi }\upsilon _{\Phi }}{\upsilon _{24}^{2}} &
		0 \\
		0 & \frac{y_{22}^{d}\upsilon _{24}\upsilon _{\Phi }}{\Lambda ^{2}} & 0 \\
		0 & 0 & \frac{y_{33}^{d}\upsilon _{\Omega }}{\Lambda }%
	\end{array}%
	\right) =\left(
	\begin{array}{ccc}
		b_{11} & b_{12} & 0 \\
		0 & b_{22} & 0 \\
		0 & 0 & b_{33}%
	\end{array}%
	\right)  \label{yd} \\
	\mathcal{Y}_{e} &=&\left(
	\begin{array}{ccc}
		\frac{4}{9}\frac{y_{11}^{d}\upsilon _{\phi }}{\upsilon _{24}^{2}} & 0 & 0 \\
		\frac{4}{9}\frac{y_{12}^{d}\upsilon _{\varphi }\upsilon _{\Phi }}{\upsilon
			_{24}^{2}} & \frac{9}{2}\frac{y_{22}^{d}\upsilon _{24}\upsilon _{\Phi }}{%
			\Lambda ^{2}} & 0 \\
		0 & 0 & \frac{y_{33}^{d}\upsilon _{\Omega }}{\Lambda }%
	\end{array}%
	\right) =\left(
	\begin{array}{ccc}
		\frac{4}{9}b_{11} & 0 & 0 \\
		\frac{4}{9}b_{12} & \frac{9}{2}b_{22} & 0 \\
		0 & 0 & b_{33}%
	\end{array}%
	\right)  \label{ye}
\end{eqnarray}%
These Yukawa matrices imply diagonal Yukawa couplings $y_{d}=b_{11}$, $%
y_{s}=b_{22}$, $y_{b}=b_{33}$, $y_{e}=\frac{4}{9}b_{11}$, $y_{\mu }=\frac{9}{%
	2}b_{22}$ and $y_{\tau }=b_{33}$ where $y_{d}$, $y_{s},$ and $y_{b}$ stand
for the eigenvalues of the down quark Yukawa matrix $\mathcal{Y}_{d}$, and $%
y_{e}$, $y_{\mu },$ and $y_{\tau }$ stand for the eigenvalues of the charged
lepton Yukawa matrix $\mathcal{Y}_{e}$. Thus, we find that for the third
family Yukawa coupling we have the well-known $b-\tau $ unification; $%
y_{\tau }=y_{b}$ which is still compatible with experimental
constraints \cite{R39}, while for the first two families -- instead
of the GJ relation -- we find alternative GUT predictions with
modified CG factors
given as%
\begin{equation}
	\frac{y_{e}}{y_{d}}=\frac{4}{9}\quad ,\quad \frac{y_{\mu }}{y_{s}}=\frac{9}{2%
	}  \label{dr}
\end{equation}%
In general, to test the validity of the GUT Yukawa couplings in a model of low-energy SUSY such as the MSSM, an accurate incorporation of SUSY threshold effects
is necessary especially in the case of large or medium $\tan \beta $ \cite%
{R33,R34,R35,R36}. However, there have been some studies showing
that the threshold corrections may be ignored when the running of
fermion masses to the GUT scale are included, see, e.g.,
\cite{R37,R38}. Moreover, to check the validity of the ratios in eq. (\ref{dr}), there are two particular
constraints for their GUT values developed in reference \cite{A12};
these are
given by%
\begin{equation}
	\frac{(1+\bar{\eta}_{l})y_{e}}{(1+\bar{\eta}_{q})y_{d}}\approx
	0.41_{-0.06}^{+0.02}\quad ,\quad \frac{(1+\bar{\eta}_{l})y_{\mu }}{(1+\bar{%
			\eta}_{q})y_{s}}\approx 4.36\pm 0.23
\end{equation}%
where $\bar{\eta}_{l}$ and $\bar{\eta}_{q}$ denote the threshold
correction parameters while the numerical values represent the
$1\sigma $ uncertainties. From these relations, one can derive the
following double ratio at the GUT scale independent of the threshold
corrections \cite{A12}
\begin{equation}
	\frac{y_{\mu }}{y_{s}}\frac{y_{d}}{y_{e}}\approx 10.7_{-0.8}^{+1.8}
	\label{rd}
\end{equation}%
As a result, we find in the present model that the ratios between
Yukawa couplings of the first two generations given in eq.
(\ref{dr}) give rise to the relation $\frac{y_{\mu
}}{y_{s}}\frac{y_{d}}{y_{e}}\simeq 10.12$ which is consistent with
the double ratio at the GUT scale given in eq. (\ref{rd}).

\begin{itemize}
	\item \textbf{Comments on proton decay}
\end{itemize}

Before we turn to the neutrino sector, we give brief comments on proton
decay which is one of the most important predictions in GUTs. It is well
known that, in the framework of the minimal SUSY $SU(5)$, the fast proton
decay comes from the contributions of the dimension four $TFF$ and dimension
five $TTTF$ baryon number violating operators\footnote{%
	For a brief review on proton decay coming from dimension four and five
	operators, see for example the appendix C of ref. \cite{B4}}. These
operators lead to proton lifetime lower than the limit provided by the
Super-Kamiokande experiment \cite{A25}. It has been shown that the $d=4$
operators can be prevented by imposing the usual $R$ symmetry like in the
case of the MSSM \cite{B18}. On the other hand, the $d=5$ operators $%
T_{i}T_{j}T_{k}F_{l}$ are generically induced via the exchange of color
triplet Higgsino \cite{B19,B20}. This issue of Higgsino-mediated proton decay%
\footnote{%
	The proton decay via dimensional five operators is mediated by the heavy
	color triplet Higgsino, and is obtained after integrating out the colored
	Higgs triplet, for more details see for example refs. \cite{A26,A27,A28}.}
is intimately connected with the so-called doublet-triplet splitting
problem---that is the problem of differentiating between the masses of the
Higgs triplets and the Higgs doublets contained in the five dimensional
Higgs of the $SU(5)$ GUT---The most effective ways proposed in the
literature to resolve this splitting problem is provided by the missing
partner (MP) and the double missing partner (DMP) mechanisms \cite%
{A29,A30,A31,A32}.

In our model, the renormalizable operators $T_{i}F_{j}F_{k}$ are
forbidden because they transform nontrivially under the $U(1)$
symmetry and thus, the proton stability at dimension 4 is
guaranteed. Moreover, the $d=4$ operators that may arise from the
coupling with the flavon fields present in the model are also
prevented by the $U(1)$ symmetry as can be checked easily from the
field assignments in tables (\ref{t1}-\ref{t3}). \newline For the $d=5$ operators, recall first that the usual tree-level Yukawa
couplings of the first and second generations are prevented by the $U(1)$
symmetry, and thus the usual operators $T_{i}T_{j}T_{k}F_{l}$ are absent in
our model. On the other hand, since the masses of quarks and charged leptons are generated from the effective operators in eqs.
(\ref{wu}) and (\ref{wd}) that involve flavon fields, the operator $TTTF$ inducing proton decay can arise -- after integrating out the color-triplet Higgs with GUT-scale mass $M_{T}$ -- from higher dimension-7 and dimension-8 operators of the form
\begin{equation}
	\frac{1}{M_{T}}T_{i}T_{j}T_{k}F_{l}\ \frac{f_{1}}{M}\frac{f_{2}}{M}(\frac{%
		f_{3}}{M})^{n}  \label{pd}
\end{equation}%
where $M$ stands for the cutoff scale $\Lambda $ or the VEV of the adjoint
Higgs, $n=0,1$ and $f_{i}$ stands for the flavon fields $\xi _{i=1,...,5}$,
$\phi $, $\varphi $, $\Phi $, and $\Omega $. It is clear that these
higher dimensional operators are consistent with the messenger content given
in appendix A considering that they are derived from the effective operators
in eqs. (\ref{wu}) and (\ref{wd}) which are themselves obtained after
integrating out a set of messenger fields required to make the model
renormalizable. As an illustration, after integrating out the
messenger fields $Y_{1}$ and $\overline{Y}_{1}$ from the
renormalizable superpotential $W_{up}^{Ren}\supset H_{5}T_{1}Y_{1}+%
\overline{Y}_{1}T_{1}\xi _{1}$ (see appendix A for the complete
superpotential), we obtain the first operator in eq. (\ref{wu})
given by $(1/\Lambda )T_{1}T_{1}H_{5}\xi _{1}$. Then, after
integrating out the colored Higgs triplet from this resulting operator and
the last operator in eq. (\ref{wd}), we obtain the dimension-7 operator%
\begin{equation}
	\frac{1}{M_{T}}\frac{1}{\Lambda ^{2}}T_{1}T_{1}T_{3}F_{2,3}\xi _{1}\Omega
	\label{nd}
\end{equation}%
Since $M_{T}$ and $\Lambda $ are both expected to be at the GUT scale, it is
straightforward to realize that the contribution of the operator (\ref{nd})
to proton decay is sufficiently suppressed. The same discussion holds for
all the allowed $T_{i}T_{j}T_{k}F_{l}$ operators generated
with a highly suppressed factors manifested by the ratios $\frac{\upsilon
	_{f_{1}}\upsilon _{f_{2}}}{M_{T}\Lambda ^{2}}$ for dimension-7 operators and
$\frac{\upsilon _{f_{1}}\upsilon _{f_{2}}\upsilon _{f_{3}}}{M_{T}\Lambda ^{3}%
}$ for dimension-8 operators.
\section{Neutrino sector}
\label{sec4}
The fermion sector of the supersymmetric $SU(5)$-GUT model is extended by
three right-handed neutrino superfields $N_{i=1,2,3}^{c}$ transforming as
gauge singlets $1_{i}$, and carrying quantum numbers under the $D_{4}\times
U(1)$ flavor group. Therefore, the light active neutrino masses are
generated through the famous type I seesaw mechanism. In our setup with the
quantum numbers of matter and Higgs superfields given in table (\ref{t1}),
five flavon superfields are required for $U(1)$ invariance in such a way
that they all couple only to the Majorana mass term $N_{i}^{c}N_{j}^{c}$
and they all carry the same $U(1)$ charge $q_{U(1)}=10$. Three of these
flavons denoted as $\rho _{1}$, $\rho _{2}$ and $\rho _{3}$ are assigned to
the $D_{4}$ singlets $1_{+,+}$, $1_{+,-}$ and $1_{-,-}$ respectively, while
the remaining two denoted as $\Gamma $ and $\digamma $ are assigned to the $%
D_{4}$ doublet $2_{0,0}$. Thus, by using the quantum numbers in tables (\ref%
{t1}) and (\ref{t2}), the superpotential invariant under the $SU(5)\times
D_{4}\times U(1)$ group is given by%
\begin{eqnarray}
	\mathcal{W}_{\nu } &=&\lambda _{1}N_{1}^{c}F_{1}H_{5}+\lambda
	_{2}N_{3,2}^{c}F_{2,3}H_{5}+\lambda _{3}N_{1}^{c}N_{1}^{c}\rho _{1}+\lambda
	_{4}N_{3,2}^{c}N_{3,2}^{c}\rho _{1}  \nonumber \\
	&&+\lambda _{5}N_{1}^{c}N_{3,2}^{c}\digamma +\lambda
	_{6}N_{1}^{c}N_{3,2}^{c}\Gamma +\lambda _{7}N_{3,2}^{c}N_{3,2}^{c}\rho
	_{2}+\lambda _{8}N_{3,2}^{c}N_{3,2}^{c}\rho _{3}  \label{wn}
\end{eqnarray}%
where $\lambda _{i=1,...,8}$ are Yukawa coupling constants. The first two
terms in $\mathcal{W}_{\nu }$ are the Dirac Yukawa terms leading to the
Dirac mass matrix $m_{D}$ while the remaining couplings give rise to the
Majorana mass matrix $m_{M}$. Our aim here is to achieve a configuration
from $\mathcal{W}_{\nu }$ that is consistent with the well-known trimaximal
mixing matrix which allows naturally for nonzero reactor angle $\theta _{13}$%
, nonmaximal atmospheric angle $\theta _{23}$ and for $\sin ^{2}\theta
_{12}\neq 1/3$. The $TM_{2}$ matrix is known to preserve the second column
of the famous TBM matrix which is ruled out by the data from reactor
neutrino experiments; nevertheless, since it is congruous with the solar and
atmospheric angles, it can still be used as a good zeroth-order
approximation. Before we develop our neutrino mass matrix $m_{\nu }$, let us
recall briefly some of the properties of the mass matrix acquired by TBM and
$TM_{2}$. For the TBM matrix, the mass matrix $m_{\nu }$ must respects the
well-known $\mu -\tau $ symmetry referring to the invariance of $m_{\nu }$
after the interchange of the $\mu $ and $\tau $ indices \cite%
{B13,B14,B15,B16,B17}, and the following
condition among the entries of $m_{\nu }$: $\left( m_{\nu }\right)
_{11}+\left( m_{\nu }\right) _{12}=\left( m_{\nu }\right) _{22}+\left(
m_{\nu }\right) _{23}$. The deviation from TBM is realized by adding small
perturbations to $m_{\nu }$ in such a way that the $\mu -\tau $ symmetry
gets broken. There are in particular two matrix perturbations that give rise
to a mass matrix with magic symmetry known to be consistent with $TM_{2}$ \cite{A13};
these two matrices are given as follows%
\begin{equation}
	\delta m_{\nu }^{1}=\left(
	\begin{array}{ccc}
		0 & 0 & \mathrm{k} \\
		0 & \mathrm{k} & 0 \\
		\mathrm{k} & 0 & 0%
	\end{array}%
	\right) \qquad ,\qquad \delta m_{\nu }^{2}=\left(
	\begin{array}{ccc}
		0 & \mathrm{k} & 0 \\
		\mathrm{k} & 0 & 0 \\
		0 & 0 & \mathrm{k}%
	\end{array}%
	\right)   \label{dm}
\end{equation}%
Now, let us use these properties in our superpotential $\mathcal{W}_{\nu }$
and derive the mass matrices of Dirac and Majorana neutrinos to calculate
the total neutrino mass matrix using the type I seesaw formula $m_{\nu
}=m_{D}m_{M}^{-1}m_{D}^{T}$. The Higgs doublet develops its VEV as usual $%
\left\langle H_{u}\right\rangle =\upsilon _{u}$\ while we assume that the
VEVs of the $D_{4}$\ breaking flavon fields point in the following directions%
\begin{equation}
	\left\langle \rho _{1}\right\rangle =\upsilon _{\rho _{1}}\quad ,\quad
	\left\langle \rho _{2}\right\rangle =\upsilon _{\rho _{2}}\quad ,\quad
	\left\langle \rho _{3}\right\rangle =\upsilon _{\rho _{3}}\quad ,\quad
	\left\langle \digamma \right\rangle =(\upsilon _{\digamma },\upsilon
	_{\digamma })^{T}\quad ,\quad \left\langle \Gamma \right\rangle =(0,\upsilon
	_{\Gamma })^{T}  \label{va}
\end{equation}%
The study of the potential which gives rise to the alignment of the flavon
doublets is discussed in appendix \ref{app4}. By using the tensor product of
$D_{4}$ irreducible representations given in eqs. (\ref{c1}) and (\ref{c2}%
), we find that the Dirac and Majorana mass matrices have the following forms%
\begin{eqnarray}
	m_{D} &=&\upsilon _{u}\left(
	\begin{array}{ccc}
		\lambda _{1} & 0 & 0 \\
		0 & \lambda _{2} & 0 \\
		0 & 0 & \lambda _{2}%
	\end{array}%
	\right) \quad ,\quad m_{M}=m_{M_{1}}+m_{M_{2}}  \nonumber \\
	\text{with \ }m_{M} &=&\left(
	\begin{array}{ccc}
		\lambda _{3}\upsilon _{\rho _{1}} & \lambda _{5}\upsilon _{\digamma } &
		\lambda _{5}\upsilon _{\digamma } \\
		\lambda _{5}\upsilon _{\digamma } & 0 & 2\lambda _{4}\upsilon _{\rho _{1}}
		\\
		\lambda _{5}\upsilon _{\digamma } & 2\lambda _{4}\upsilon _{\rho _{1}} & 0%
	\end{array}%
	\right) +\left(
	\begin{array}{ccc}
		0 & 0 & \lambda _{6}\upsilon _{\Gamma } \\
		0 & \lambda _{7}\upsilon _{\rho _{2}}-\lambda _{8}\upsilon _{\rho _{3}} & 0
		\\
		\lambda _{6}\upsilon _{\Gamma } & 0 & \lambda _{7}\upsilon _{\rho
			_{2}}+\lambda _{8}\upsilon _{\rho _{3}}%
	\end{array}%
	\right)
\end{eqnarray}%
The Majorana mass matrix is decomposed in terms of two matrices to show that
the TBM conditions and its deviation to the $TM_{2}$ are obtained from $%
m_{M_{1}}$ and $m_{M_{2}}$, respectively. Accordingly, the $\mu -\tau $ symmetry and the
condition $\left( m_{\nu }\right) _{11}+\left( m_{\nu }\right) _{12}=\left(
m_{\nu }\right) _{22}+\left( m_{\nu }\right) _{23}$ require the imposition
of the following assumptions on $m_{D}$\ and $m_{M_{1}}$
\begin{equation}
	\lambda _{1}=\lambda _{2}\qquad \text{and}\qquad \lambda _{3}\upsilon _{\rho
		_{1}}+\lambda _{5}\upsilon _{\digamma }=2\lambda _{4}\upsilon _{\rho _{1}} \label{as1}
\end{equation}%
while the deviation from TBM to $TM_{2}$ requires a mass matrix with the
magic symmetry which is conceivable by the imposition of the following
assumption on $m_{M_{2}}$%
\begin{equation}
	\lambda _{8}\upsilon _{\rho _{3}}=-\lambda _{7}\upsilon _{\rho _{2}}=\lambda
	_{6}\upsilon _{\Gamma }/2 \label{as2}
\end{equation}%
leading to the form of the matrix perturbation $\delta m_{\nu }^{1}$ in (%
\ref{dm}). The plausibility of these assumptions is discussed in appendix %
\ref{app4}. To simplify the parametrization of the total neutrino mass
matrix and later the expressions of neutrino masses as well as the mixing
angles, we parametrize the Majorana mass matrix as follows%
\begin{equation}
	m_{M}=M_{R}\left(
	\begin{array}{ccc}
		a & b & b+\mathrm{k} \\
		b & \mathrm{k} & c \\
		b+\mathrm{k} & c & 0%
	\end{array}%
	\right)   \label{md}
\end{equation}%
where $M_{R}$ is the mass scale of the heavy RH Majorana neutrinos and $a=%
\frac{\lambda _{3}\upsilon _{\rho _{1}}}{M_{R}}$, $b=\frac{\lambda
	_{5}\upsilon _{\digamma }}{M_{R}}$, $c=\frac{2\lambda _{4}\upsilon _{\rho
		_{1}}}{M_{R}}$ and $\mathrm{k}=\frac{\lambda _{6}\upsilon _{\Gamma }}{M_{R}}$%
. The usual canonical seesaw formula $m_{\nu }=m_{D}m_{M}^{-1}m_{D}^{T}$
yields the total neutrino mass matrix
\begin{equation}
	m_{\nu }=\frac{m_{0}}{P}\left(
	\begin{array}{ccc}
		-\left( a+b\right) ^{2} & \left( a+b\right) \left( b+\mathrm{k}\right)  &
		b^{2}-\mathrm{k}^{2}-b\left( \mathrm{k}-a\right)  \\
		\left( a+b\right) \left( b+\mathrm{k}\right)  & -\left( b+\mathrm{k}\right)
		^{2} & -a^{2}-ab+b^{2}+\mathrm{k}b \\
		b^{2}-\mathrm{k}^{2}-b\left( \mathrm{k}-a\right)  & -a^{2}-ab+b^{2}+\mathrm{k%
		}b & a\mathrm{k}-b^{2}%
	\end{array}%
	\right)   \label{mn}
\end{equation}%
where $m_{0}=\frac{\left( \lambda _{1}\upsilon _{u}\right) ^{2}}{M_{R}}$ and
$P=\left( a+2b+\mathrm{k}\right) \left( a\mathrm{k}-a^{2}+b^{2}-\mathrm{k}%
^{2}\right) $. It is clear to verify that in the limit where $\mathrm{k}%
\rightarrow 0$, this matrix obeys the $\mu -\tau $ symmetry, and thus it is diagonalized by the TBM matrix which
is $CP$ conserving and predicts $\theta _{13}=0$ and $\theta _{23}=\pi /4$.
Therefore, the presence of $\mathrm{k}$ is necessary to break the $\mu -\tau
$ symmetry and produce a small deviation from the TBM pattern as mentioned
above. Without loss of generality, only the parameter $\mathrm{k}$ is taken
to be complex -- $\mathrm{k}\rightarrow \left\vert \mathrm{k}\right\vert
e^{i\phi _{k}}$ where $\phi _{k}$ is a $CP$ violating phase -- which is
sufficient to ensure $CP$ violation in the lepton sector. On the other hand,
the matrix $m_{\nu }$\ enjoys the magic symmetry property which refers to
the equality of the sum of each row and the sum of each column in the
neutrino mass matrix. This property implies that the neutrino
matrix is diagonalized by the well-known trimaximal mixing matrix $\mathcal{U%
}_{TM_{2}}$ so that $m_{\nu }^{\mathrm{diag}}=\mathcal{U}_{TM_{2}}^{\dagger
}m_{\nu }\mathcal{U}_{TM_{2}}$\ with%
\begin{equation}
	\mathcal{U}_{TM_{2}}=\left(
	\begin{array}{ccc}
		\sqrt{\frac{2}{3}}\cos \theta  & \frac{1}{\sqrt{3}} & \sqrt{\frac{2}{3}}\sin
		\theta e^{-i\sigma } \\
		-\frac{\cos \theta }{\sqrt{6}}-\frac{\sin \theta }{\sqrt{2}}e^{i\sigma } &
		\frac{1}{\sqrt{3}} & \frac{\cos \theta }{\sqrt{2}}-\frac{\sin \theta }{\sqrt{%
				6}}e^{-i\sigma } \\
		-\frac{\cos \theta }{\sqrt{6}}+\frac{\sin \theta }{\sqrt{2}}e^{i\sigma } &
		\frac{1}{\sqrt{3}} & -\frac{\cos \theta }{\sqrt{2}}-\frac{\sin \theta }{%
			\sqrt{6}}e^{-i\sigma }%
	\end{array}%
	\right)   \label{3-11}
\end{equation}%
The full mixing matrix is given by $\mathcal{U}_{\nu }=\mathcal{U}_{TM_{2}}%
\mathcal{U}_{P}$ where
$\mathcal{U}_{P}=\mathrm{diag}(1,e^{i\frac{\alpha
		_{21}}{2}},e^{i\frac{\alpha _{31}}{2}})$ is a diagonal matrix that
contains the Majorana phases $\alpha _{21}$ and $\alpha _{31}$. The
parameters $\theta $ and $\sigma $ are respectively an arbitrary
angle and a phase which will be related to the neutrino oscillation
parameters; the
observed neutrino mixing angles $\theta _{ij}$ and the Dirac $CP$ phase $%
\delta _{CP}$. The diagonalization of the neutrino matrix \ref{mn}
by $\mathcal{U}_{TM_{2}}$ induces relations between our model
parameters and
the trimaximal mixing parameters $\sigma $ and $\theta $, we find%
\begin{equation}
	\tan 2\theta =\frac{\sqrt{3}\left\vert \mathrm{k}\right\vert \sqrt{b^{2}\cos
			^{2}\phi _{\mathrm{k}}+a^{2}\sin ^{2}\phi _{k}}}{2ab-b\left\vert \mathrm{k}%
		\right\vert \cos \phi _{k}}\quad ,\quad \tan \sigma =\frac{-a}{b}\tan \phi
	_{k}
\end{equation}%
As a result, the eigenmasses of $m_{\nu }$ are as follows%
\begin{equation}
	\begin{array}{c}
		\left\vert m_{1}\right\vert =\frac{m_{0}}{\sqrt{(a-b)^{2}-\left\vert \mathrm{%
					k}\right\vert (a-b)\cos \phi _{k}+(\left\vert \mathrm{k}\right\vert ^{2}/4)}}%
		\quad ,\quad \left\vert m_{2}\right\vert =\frac{m_{0}}{\sqrt{%
				(a+2b)^{2}+2\left\vert \mathrm{k}\right\vert (a+2b)\cos \phi _{k}+\left\vert
				\mathrm{k}\right\vert ^{2}}} \\
		\left\vert m_{3}\right\vert =\frac{m_{0}}{\sqrt{(a+b)^{2}-\left\vert \mathrm{%
					k}\right\vert (a+b)\cos \phi _{k}+(\left\vert \mathrm{k}\right\vert ^{2}/4)}}%
	\end{array}
	\label{mas}
\end{equation}%
where the denominators of these masses corresponds to ratios of the
right-handed neutrino masses and their mass scale $M_{R}$%
\begin{eqnarray}
	\frac{\left\vert M_{1}\right\vert }{M_{R}} &=&\sqrt{(a-b)^{2}-\left\vert
		\mathrm{k}\right\vert (a-b)\cos \phi _{k}+(\left\vert \mathrm{k}\right\vert
		^{2}/4)}  \notag \\
	\frac{\left\vert M_{2}\right\vert }{M_{R}} &=&\sqrt{(a+2b)^{2}+2\left\vert
		\mathrm{k}\right\vert (a+2b)\cos \phi _{k}+\left\vert \mathrm{k}\right\vert
		^{2}}  \label{Mi} \\
	\frac{\left\vert M_{3}\right\vert }{M_{R}} &=&\sqrt{(a+b)^{2}-\left\vert
		\mathrm{k}\right\vert (a+b)\cos \phi _{k}+(\left\vert \mathrm{k}\right\vert
		^{2}/4)}  \notag
\end{eqnarray}%
Regarding the mixing angles, it is well-known that the total lepton mixing
matrix is derived from the product between two matrices; $U_{PMNS}=\mathcal{U%
}_{l}^{\dagger }\mathcal{U}_{\nu }$ where $\mathcal{U}_{l}$ is the matrix
that diagonalizes the charged lepton mass matrix while $\mathcal{U}_{TM_{2}}$
is as described above. In the present model however, it is easy to check
that the charged lepton mixing angles $\theta _{ij}^{l}$ derived from the
diagonalization of $m_{l}=\upsilon _{d}\mathcal{Y}_{e}$---where $\mathcal{Y}%
_{e}$ is given in eq. (\ref{ye})---are all equals to zero; thus, they do not
affect the neutrino mixing angles derived from $\mathcal{U}_{\nu }$.
Therefore, by using the PDG standard parametrization of the PMNS matrix \cite%
{C1},$\allowbreak $ the reactor, solar and atmospheric angles are expressed
as%
\begin{equation}
	\sin ^{2}\theta _{13}=\frac{2}{3}\sin ^{2}\theta ~,~\sin ^{2}\theta _{12}=%
	\frac{1}{3-2\sin ^{2}\theta }~,~\sin ^{2}\theta _{23}=\frac{1}{2}-\frac{%
		3\sin 2\theta }{2\sqrt{3}(3-2\sin ^{2}\theta )}\cos \sigma \text{.}
	\label{mix}
\end{equation}%
Before we perform a numerical analysis of neutrino masses and model
parameters, we should notice that the above masses and mixing
parameters are valid at the GUT scale. Therefore, to match these
parameters with the experimental values (of the mixing angles, the
$CP$ phase and the mass squared differences), their evolution from
the GUT scale to low energy must be carried out. However, although
the final values are model dependent, it was illustrated in ref.
\cite{A24} that for SUSY models, if $\tan \beta $ is small, the
RG-induced effects on the above parameters are controllable and can
be safely neglected. Accordingly, since the type I seesaw mechanism
is related to physics at very high energy scales, we can work in a
scenario with small neutrino Yukawa couplings in such a way that
their contribution to the RG evolution can be neglected \cite{A12}.
\section{Numerical analysis and results}
\label{sec5} In this section, we carry out a detailed numerical
analysis for both charged fermion and neutrino sectors. For the
charged fermion sector, we fix the values of the model parameters in
order to reproduce the observed fermion Yukawa couplings and the CKM
mixing parameters at the GUT scale within $1\sigma $ ranges. As for
the neutrino sector, we constrain our model parameters using the $3\sigma $ allowed range of the neutrino oscillation parameters. We also use constraints from non-oscillatory experiments to make predictions concerning the physical observables $m_{\beta \beta}$, $m_{\beta}$, and $\sum m_{i}$.
\subsection{Numerical fits for charged fermion sector}
Our model predicts the Yukawa couplings and mixing parameters at the
GUT scale which we assumed to be also the flavor symmetry breaking
scale. To compare the obtained spectrum of our model with the data
extrapolated at the GUT scale, the experimental values must run up
to the GUT scale taking into account the SUSY parameters; $\tan
\beta $ and SUSY threshold correction effects. Such an analysis has
been performed in ref. \cite{A12}, where the extracted CKM
parameters and all Yukawa couplings at
the GUT scale for $\tan \beta =5$ and $\tan \beta =10$ with $M_{SUSY}=1$ $%
\mathrm{TeV}$ and $\eta _{b}=-0.2437$ are given in table (3) of ref. \cite{B1}%
. In our model, recall\ that the mass matrices of the charged fermions are
generally expressed as%
\begin{equation}
	m_{up}=\upsilon _{u}\mathcal{Y}_{up}\qquad ,\qquad m_{down}=\upsilon _{d}%
	\mathcal{Y}_{d}\qquad ,\qquad m_{lep}=\upsilon _{d}\mathcal{Y}_{e}
\end{equation}%
where $\mathcal{Y}_{up}$, $\mathcal{Y}_{d}$ and $\mathcal{Y}_{e}$
refer to the Yukawa matrices of up quarks, down quarks and charged
leptons obtained in eqs. (\ref{yu}), (\ref{yd}) and (\ref{ye}),
respectively. Recall also that the ratio between the Higgs VEVs
$\upsilon _{u}$ and $\upsilon _{d}$ is defined as $\tan \beta
=\frac{\upsilon _{u}}{\upsilon _{d}}$ while the SM Higgs VEV is
$\upsilon =\sqrt{\upsilon _{u}^{2}+\upsilon _{d}^{2}}=174$
$\mathrm{GeV}$, we have%
\begin{equation}
	m_{up}=\upsilon \sin \beta \mathcal{Y}_{up}\qquad ,\qquad m_{down}=\upsilon
	\cos \beta \mathcal{Y}_{d}\qquad ,\qquad m_{lep}=\upsilon \cos \beta
	\mathcal{Y}_{e}
\end{equation}%
We assume in our calculations that all Yukawa coupling constants are
real except for $y_{16}$ which is complex and leads to the complex
nature of the parameter $b_{12}=\left\vert b_{12}\right\vert
e^{i\epsilon }$, and subsequently to the $CP$ violation in the quark
sector. Moreover, the Yukawa matrices in eqs. (\ref{yu}) and
(\ref{yd}) involve the following independent parameters [$
a_{11}$, $a_{12}$, $a_{13}$, $a_{22}$, $a_{23}$, $a_{33}$, $b_{11}$, $b_{12}$, $\left\vert
b_{22}\right\vert$, $b_{33}$ and $\epsilon $], which we
need to fix in order to perform our numerical analysis at the GUT
scale.
\begin{table}[h]
	\centering
	\begin{tabular}{|l|l||ll}
		\hline
		Model parameters & Values & Model parameters & \multicolumn{1}{|l|}{Values}
		\\ \hline
		$a_{11}$ & $0.1189\times 10^{-4}$ & $b_{11}$ & \multicolumn{1}{|l|}{$%
			0.44929\times 10^{-4}$} \\ \hline
		$a_{12}$ & $0.1530\times 10^{-2}$ & $\left\vert b_{12}\right\vert $ &
		\multicolumn{1}{|l|}{$0.240406\times 10^{-3}$} \\ \hline
		$a_{13}$ & $0.2487\times 10^{-2}$ & $b_{22}$ & \multicolumn{1}{|l|}{$%
			0.9545\times 10^{-2}$} \\ \hline
		$a_{22}$ & $0.2076\times 10^{-2}$ & $b_{33}$ & \multicolumn{1}{|l|}{$0.0705$}
		\\ \hline
		$a_{23}$ & $0.01857$ & $\epsilon $ & \multicolumn{1}{|l|}{$\frac{\pi }{3}$}
		\\ \hline
		$a_{33}$ & $0.5218$ &  &  \\ \cline{1-2}
	\end{tabular}%
	\caption{Parameters of the quark and charged leptons Yukawa matrices at the
		GUT scale with $\tan \protect\beta =10$.}
	\label{t6}
\end{table}
Next, we fix these input parameters for two cases: $\tan \beta =10$ and $%
\tan \beta =5$. For $\tan \beta =10$, the numerical values are
reported in table (\ref{t6}). The values reported in this table are
fixed respecting the fact that the magnitude of the flavon VEVs are
smaller than the flavor symmetry breaking scale; $\upsilon
_{flavons}<M_{GUT}$, while we fix the phase $\epsilon $ to the
value $\frac{\pi }{3}$ which yields the correct
experimental fit of the $CP$-violating Dirac phase of the quark sector.%
The above estimates concerning the input parameters produce the
values of the physical quantities -- namely the quark mixing angles,
the Yukawa couplings and the $CP$ phase -- at the GUT scale; these
numerical values are as reported in table (\ref{t7}).
\begin{table}[h]
	\centering%
	\begin{tabular}{|l|l||llll}
		\hline
		Observables & values & Observables & \multicolumn{1}{|l}{values} & 
		\multicolumn{1}{||l}{Observables} & \multicolumn{1}{|l|}{values} \\ \hline
		$\theta _{12}^{q}/^{\circ }$ & $13.0433$ & $y_{u}$ & \multicolumn{1}{|l}{$%
			2.88221\times 10^{-6}$} & \multicolumn{1}{||l}{$y_{d}$} & 
		\multicolumn{1}{|l|}{$4.33494\times 10^{-6}$} \\ \hline
		$\theta _{13}^{q}/^{\circ }$ & $0.18015$ & $y_{c}$ & \multicolumn{1}{|l}{$%
			1.40921\times 10^{-3}$} & \multicolumn{1}{||l}{$y_{s}$} & 
		\multicolumn{1}{|l|}{$9.79486\times 10^{-5}$} \\ \hline
		$\theta _{23}^{q}/^{\circ }$ & $2.05405$ & $y_{t}$ & \multicolumn{1}{|l}{$%
			0.51988$} & \multicolumn{1}{||l}{$y_{b}$} & \multicolumn{1}{|l|}{$%
			7.01501\times 10^{-3}$} \\ \hline
		$\delta _{CP}^{q}/^{\circ }$ & $69.166$ &  &  &  &  \\ \cline{1-2}
	\end{tabular}%
	\caption{The predictions for the Yukawa eigenvalues, the mixing angles and
		the $CP$ phase of the quark sector for $\tan \protect\beta =10$.}
	\label{t7}
\end{table}
We repeat the same numerical fit for $\tan \beta =5$, where the
input parameters and the output for the physical parameters are
reported in tables (\ref{t8}) and (\ref{t9}), respectively.
\begin{table}
	\centering
	\begin{tabular}{|l|l||ll}
		\hline
		Model parameters & Values & Model parameters & \multicolumn{1}{|l|}{Values}
		\\ \hline
		$a_{11}$ & $0.12397\times 10^{-4}$ & $b_{11}$ & \multicolumn{1}{|l|}{$%
			0.226238\times 10^{-4}$} \\ \hline
		$a_{12}$ & $0.15866\times 10^{-3}$ & $\left\vert b_{12}\right\vert $ &
		\multicolumn{1}{|l|}{$0.11827\times 10^{-3}$} \\ \hline
		$a_{13}$ & $0.25912\times 10^{-2}$ & $b_{22}$ & \multicolumn{1}{|l|}{$%
			0.47008\times 10^{-3}$} \\ \hline
		$a_{22}$ & $0.21459\times 10^{-2}$ & $b_{33}$ & \multicolumn{1}{|l|}{$%
			0.036597$} \\ \hline
		$a_{23}$ & $0.019344$ & $\epsilon $ & \multicolumn{1}{|l|}{$\frac{\pi }{3}$}
		\\ \hline
		$a_{33}$ & $0.5435$ &  &  \\ \cline{1-2}
	\end{tabular}%
	\caption{Parameters of the quark and charged leptons Yukawa matrices
		at the GUT scale with $\tan \protect\beta =5$.} \label{t8}
\end{table}
\begin{table}[tbp]
	\centering%
	\begin{tabular}{|l|l||llll}
		\hline
		Observables & values & Observables & \multicolumn{1}{|l}{values} & 
		\multicolumn{1}{||l}{Observables} & \multicolumn{1}{|l|}{values} \\ \hline
		$\theta _{12}/^{\circ }$ & $13.0295$ & $y_{u}$ & \multicolumn{1}{|l}{$%
			2.92027\times 10^{-6}$} & \multicolumn{1}{||l}{$y_{d}$} & 
		\multicolumn{1}{|l|}{$4.16836\times 10^{-6}$} \\ \hline
		$\theta _{13}/^{\circ }$ & $0.18019$ & $y_{c}$ & \multicolumn{1}{|l}{$%
			1.42946\times 10^{-3}$} & \multicolumn{1}{||l}{$y_{s}$} & 
		\multicolumn{1}{|l|}{$9.21045\times 10^{-5}$} \\ \hline
		$\theta _{23}/^{\circ }$ & $2.05418$ & $y_{t}$ & \multicolumn{1}{|l}{$0.53331
			$} & \multicolumn{1}{||l}{$y_{b}$} & \multicolumn{1}{|l|}{$6.95358\times
			10^{-3}$} \\ \hline
		$\delta _{CP}^{q}/^{\circ }$ & $69.1801$ &  &  &  &  \\ \cline{1-2}
	\end{tabular}%
	\caption{The predictions for the Yukawa eigenvalues, the mixing angles and
		the $CP$ phase of the quark sector for $\tan \protect\beta =5$}
	\label{t9}
\end{table}
This fit has been performed using the Mixing Parameter Tools package \cite%
{A24}. The obtained values are in a good agreement with the GUT
scale data for both $\tan \beta =5$ and $\tan \beta =10$
\cite{A12,B1}.
\subsection{Neutrino phenomenology}
The fact that the neutrino mass ordering remains unknown requires
the investigation of the two possible options: either $\Delta
m_{31}^{2}$ $>0$ referred to as normal mass hierarchy or $\Delta
m_{32}^{2}$ $<0$ known as the inverted mass Hierarchy (IH). In the
latter case that implies $m_{3}<m_{2}<m_{1}$, it is easy to deduce
from the first relation in eq. (\ref{mix}) as well as the $3\sigma $
region of the reactor angle from ref. \cite{R3} that the parameter $\theta $ lies in the interval $%
0.1763\leq \theta \leq 0.1920$.
On the other hand, by requiring the
values of the mass-squared differences $\Delta m_{ij}^{2}$ within
their $3\sigma $ experimental ranges and using the eigenmasses in eq. (%
\ref{mas}) as well as the constraint on the sum of
neutrino masses from cosmological observations $\sum
m_{i}<0.12$ $\mathrm{eV}$ \cite{A23}, we find that $\theta $ lies in the interval $0.398\lesssim \theta
\lesssim 0.579$ which implies that both $\sin ^{2}\theta _{13}$ and
$\sin ^{2}\theta _{12}$ fall far outside their $3\sigma $
experimental range. For this reason, the IH scheme is excluded in
our model.

\begin{figure}
	\begin{center}
		\includegraphics[width=.45\textwidth]{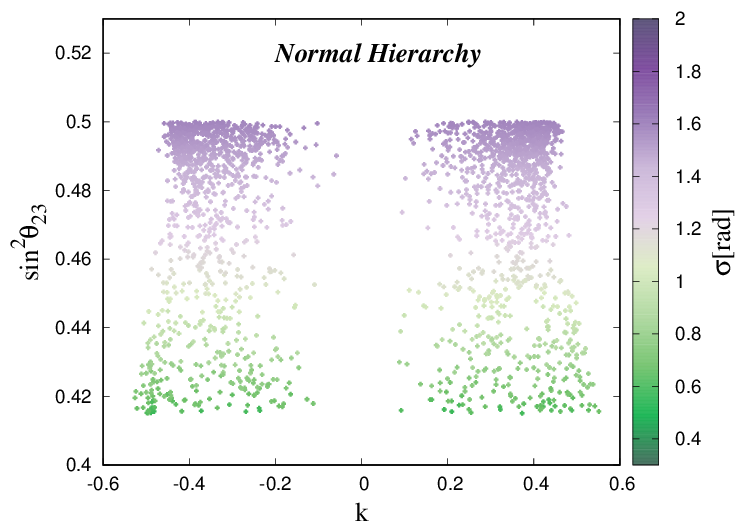} \includegraphics[width=.45%
		\textwidth]{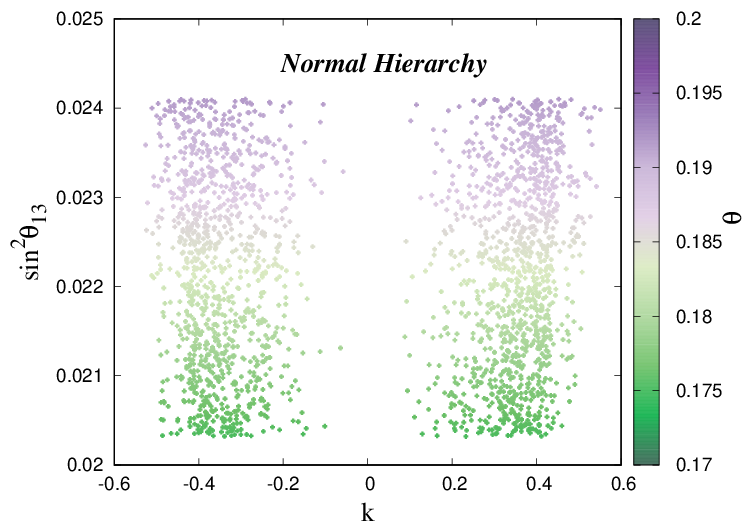} \includegraphics[width=.45\textwidth]{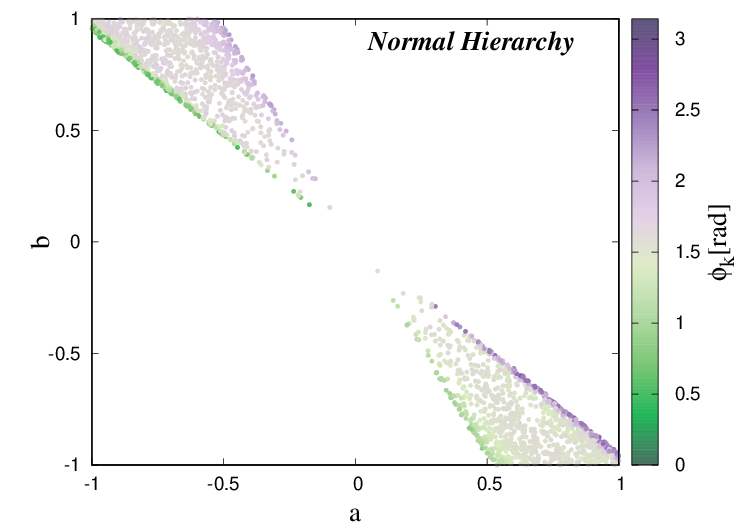}
	\end{center}
	\vspace{-2em}
	\caption{The allowed points of the trimaximal mixing parameters $\protect%
		\sigma $ (top left panel) and $\protect\theta $ (top right panel) projected
		on the planes $\left( \sin ^{2}\protect\theta _{23},\mathrm{k}\right) $ and $%
		\left( \sin ^{2}\protect\theta _{13},\mathrm{k}\right) $ respectively. The
		bottom panel shows the allowed points in the parameter space of ($a$, $b$)
		with the color code indicating the phase $\protect\phi _{k}$.}
	\label{f1}
\end{figure}

As regards to the NH scheme, we rewrite the masses $m_{2}$ and $m_{3}$%
in terms of the lightest neutrino mass $m_{1}$ and the mass
squared differences as $m_{2}=\sqrt{m_{1}^{2}+\Delta m_{21}^{2}}$ and\ $%
m_{3}=\sqrt{m_{1}^{2}+\Delta m_{31}^{2}}$. Moreover, by using eqs. (\ref{mas}%
) and (\ref{mix}), these masses can also be expressed as a function
of the free parameters $a$, $b$, \textrm{k} and $\phi _{k}$ to which
ascribe the smallness of neutrino masses. Thus, we allow $a$, $b$
and $\mathrm{k}$ to vary in the range $[-1,1]$ while we allow $\phi
_{k}$ to vary
in the range $[0,\pi ]$. In figure (\ref{f1}), the trimaximal mixing parameters $%
\sigma $ (top left panel) and $\theta $ (top right panel) are
projected on the planes $\left( \sin ^{2}\theta
_{23},\mathrm{k}\right) $ and $\left( \sin ^{2}\theta
_{13},\mathrm{k}\right) $ respectively. As inputs, the angle $\theta
$ is allowed to vary in the range $\left[ 0, \pi /2\right]
$ while the phase $\sigma $ is randomly varied in the range $\left[
0, 2\pi \right] $. From the left panel of figure (\ref{f1}),
we observe from the scattered points that for the atmospheric angle
only the lower octant ($\sin ^{2}\theta _{23}<0.5$) is allowed.
Therefore, an important prediction of the current model is that it
excludes the maximal as
well as the higher octant of the atmospheric angle. Moreover, the $3\sigma $%
allowed intervals of the oscillation parameters restrict the range of $%
\sigma $ as well as the range of the parameter \textrm{k}
\begin{equation}
	0.57565\lesssim \sigma \lesssim 1.57073\qquad ,\qquad -0.52597\lesssim
	\mathrm{k}\lesssim 0.55115  \label{sk}
\end{equation}
Notice by the way that the parameter $\mathrm{k}$ is responsible for
the deviation from the TBM values of the mixing angles. This
deviation is encoded in the parameter $\theta $ which is easily seen
when we set $\theta \rightarrow 0$ in eq. (\ref{mix}) resulting to
restore the TBM values. From the top right panel of figure (\ref{f1}),
we find that the range of $\theta $ is also restricted to
$0.17548\lesssim \theta \lesssim 0.19129$ while the range of the
reactor angle remains almost unchanged compared to its $3\sigma $ allowed range.
In figure (\ref{f1}), the bottom panel shows the correlation between
the parameters $a$ and $b$ with the color code showing the phase
$\phi _{k}$. By
taking into account the $3\sigma $ experimental ranges of $\Delta m_{ij}^{2}$%
, $\sin \theta _{ij}$ and $\delta _{CP}$ from the most recent global fit by
NuFIT collaboration \cite{R3}, and the current cosmological upper bound on
the sum of the three light neutrino masses given by $\sum m_{i}<0.12$ $%
\mathrm{eV}$, we find that the range of the phase $\phi _{k}$ gets more
restricted compared to its input range; $0.55275\lesssim \phi _{k}\lesssim
2.56095$. Regarding the Dirac $CP$ phase $\delta _{CP}$, the results
reported by the T2K long-baseline experiment showed strong hints for $CP$
violation in neutrino oscillations while $CP$ conservation is disfavored at $%
2\sigma $ level \cite{R6}. One approach to estimate the magnitude of
$\delta _{CP}$ is by means of the Jarlskog invariant parameter
defined as
$J_{CP}=Im(\mathcal{U}_{e1}\mathcal{U}_{\mu 1}^{\ast }\mathcal{U}%
_{\mu 2}\mathcal{U}_{e2}^{\ast })$. In the PDG standard parametrization,
this parameter is exhibited in terms of the three mixing angles and the
Dirac $CP$ phase as follows \cite{C1}%
\begin{equation}
	J_{CP}=\frac{1}{8}\sin 2\theta _{12}\sin 2\theta _{13}\sin 2\theta _{23}\cos
	\theta _{13}\sin \delta _{CP}  \label{jcp}
\end{equation}%
while in the case of the trimaximal mixing, it takes a simpler form given by%
$J_{CP}^{TM}=\left( 1/6\sqrt{3}\right) \sin 2\theta \sin \sigma $%
. By matching $J_{CP}^{TM}$ with eq. (\ref{jcp}), we find a
correlation between the Dirac $CP$ phase, the arbitrary phase
$\sigma $, and the
atmospheric angle%
\begin{equation}
	\sin 2\theta _{23}\sin \delta _{CP}=\sin \sigma  \label{ds}
\end{equation}%
Taking into account the fact that atmospheric angle is well determined
experimentally as well as the fact that the range of $\sigma $ given in eq. (%
\ref{sk}) excludes the exact value of $n\pi $ with $n$ can be any
integer, it is easy to deduce analytically that the $CP$ conserving
values of $\delta _{CP}$ are not allowed which implies that the
present model admits only the $CP$ violating values of $\delta
_{CP}$.
\begin{itemize}
	\item Neutrino masses from non-oscillatory experiments
\end{itemize}

Constraining the absolute neutrino mass scale is one of the most important
purposes of the forthcoming neutrino experiments. This scale can be probed
by various non-oscillatory neutrino experiments. Cosmological observations
are in particular a powerful tool to probe the total sum of neutrino masses.
Indeed, in the framework of $\Lambda $CDM model with three massive active
neutrinos, the latest Planck data combined with baryon acoustic oscillations
(BAO) measurements provided an upper bound on the sum of neutrino masses of $%
\sum m_{i}<0.12$ $\mathrm{eV}$ \cite{A23}; see also ref. \cite{D2} for a comprehensive analysis of the changes in the upper bounds of $\sum m_{i}$ after taking into account neutrino oscillation data. Another way to probe this scale
is through direct neutrino mass determination where the study of the
electron energy spectrum near its endpoint region is up to date the most
sensitive method to determine the electron antineutrino mass. The effective
electron neutrino mass is defined in terms of the three neutrino mass
eigenvalues $m_{i}$ and the flavor mixing parameters $U_{ei}$ as $m_{\beta
}=\left( \sum_{i}\left\vert U_{ei}\right\vert ^{2}m_{i}^{2}\right) ^{1/2}$.
Currently, the most valid bounds on $m_{\beta }$ are presented by the KATRIN
experiment which provides an upper limit on the electron antineutrino mass
of $1.1$ $\mathrm{eV}$ \cite{C2} and eventually aims at a sensitivity of $%
0.2 $ $\mathrm{eV}$ \cite{C3}.
\begin{figure}[h]
	\begin{center}
		\includegraphics[width=.44\textwidth]{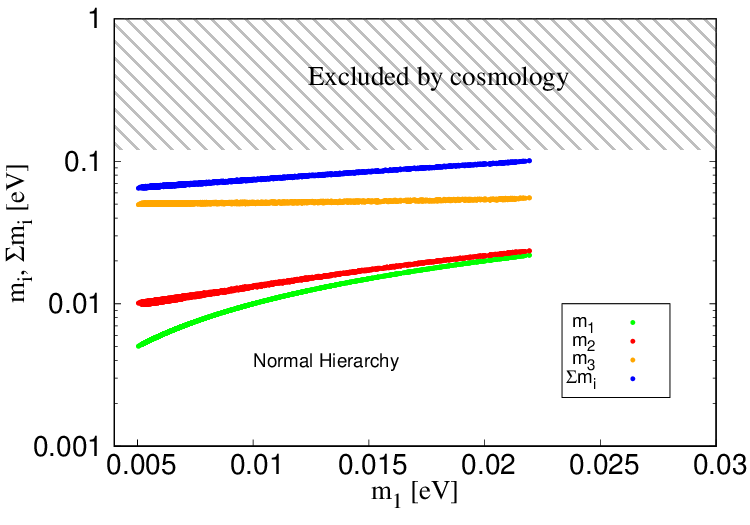} \includegraphics[width=.44%
		\textwidth]{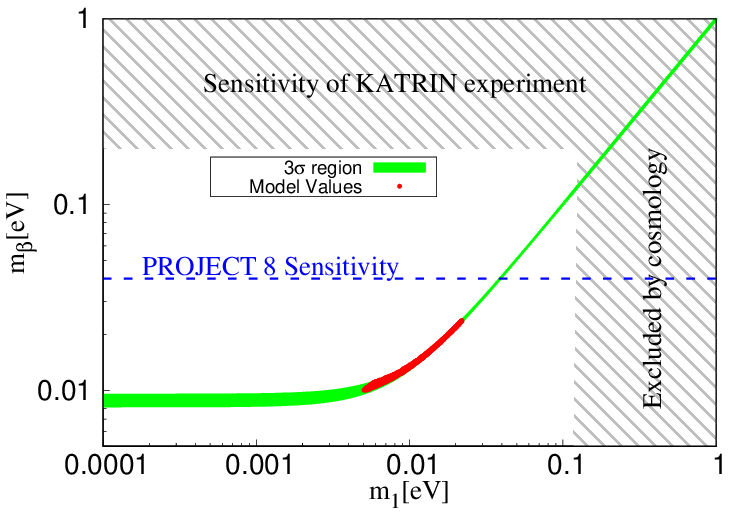}
	\end{center}
	\vspace{-2em}
	\caption{Left: Prediction for the absolute neutrino masses and their sum $%
		\sum m_{i}$ as a function of $m_{1}$. Right: $m_{\protect\beta }$ as a
		function of $m_{1}$ where the gray dashed vertical region is disfavored by
		Planck+BAO while the gray dashed horizontal region is the limit on $m_{%
			\protect\beta }$ from KATRIN collaboration.}
	\label{f3}
\end{figure}
Using the upper limit $\sum m_{i}<0.12$ $\mathrm{eV}$ and the
neutrino oscillation parameters ($\theta _{ij}$ and $\Delta
m_{ij}^{2}$) within their currently allowed $3\sigma $ ranges as
well as the restricted interval of our model parameters given in the
previous subsection, we show
in the left panel of figure (\ref{f3}) the three neutrino masses $m_{1}$, $%
m_{2}$ and $\ m_{3}$ given in eq. (\ref{mas}), and their sum $\sum
m_{i}$ as a function of the lightest neutrino mass $m_{1}$. We find
that our model
predicts the following ranges%
\begin{eqnarray}
	0.064759 &\lesssim &\sum m_{i}\left( \mathrm{eV}\right) \lesssim
	0.100929\quad ,\quad 0.005033\lesssim m_{1}\left( \mathrm{eV}\right)
	\lesssim 0.021934  \notag \\
	0.009851 &\lesssim &m_{2}\left( \mathrm{eV}\right) \lesssim 0.023530\quad ,%
	\text{\quad }0.049614\lesssim m_{3}\left( \mathrm{eV}\right) \lesssim
	0.055478  \label{m1}
\end{eqnarray}%
As a result, the predicted values of $\sum m_{i}$ around the lower bound $%
\sim 0.064759$ $\mathrm{eV}$ are consistent with normal mass hierarchy
which requires\footnote{%
	This bound is obtained by taking the best fit values of the mass squared
	differences $\Delta m_{21}^{2}$ and $\Delta m_{31}^{2}$ from ref. \cite{R3}
	with any value of the lightest neutrino mass $m_{1}$ obtained in eq. (\ref%
	{m1}).} $\sum m_{i}\left( \mathrm{eV}\right) \gtrsim 0.065431$.%
This lower bound of $\sum m_{i}$ may be achieved in the forthcoming
experiments with further cosmological data such as CORE+BAO aiming
to reach a $0.062$ $\mathrm{eV}$ sensitivity on
the sum of the three active neutrino masses \cite{C4}.%
\newline
In the right panel of figure (\ref{f3}), we show the correlation between $%
m_{\beta }$ and the lightest neutrino mass $m_{1}$. The orange
region is achieved by varying all the input parameters ($\Delta
m_{ij}^{2}$, $a$, $b$, $\left\vert \mathrm{k}\right\vert $ and $\phi
_{k}$) in their $3\sigma $ ranges while the red points stands for
our model prediction. We
find that the effective electron neutrino mass is given by%
\begin{equation}
	0.0100158\lesssim m_{\beta }\left( \mathrm{eV}\right) \lesssim
	0.023765
\end{equation}%
It is clear that our predictions for $m_{\beta }$ are too small when
compared to the anticipated future $\beta $-decay experiments
sensitivities such as KATRIN ($\sim 0.2$ \textrm{eV}) \cite{C3}, HOLMES ($%
\sim 0.1$ \textrm{eV}) \cite{C5}, and Project 8 ($\sim 0.04$ \textrm{eV})
\cite{C6}. If the actual electron neutrino mass would be measured by one of
these experiments the neutrino sector of the present model will be ruled
out. Otherwise, the obtained values could be probed by new experimental
projects that must aim to reach improved sensitivities around $0.01$ $%
\mathrm{eV}$.

Another possible portal to probe the scale of neutrino masses comes from
experiments exploring the nature of neutrinos which is also one of the
present objectives in the field of neutrino physics. Up to now, the probe of
the Majorana nature of neutrinos is available only through $0\nu \beta \beta
$ decay. This is a process that violates lepton number $\emph{L}$ by two
units, and since there are no SM interactions that violates $\emph{L}$, the
discovery of $0\nu \beta \beta $ would have interesting implications for
model building beyond the SM such as the existence of a new mechanism for
mass generation compared to the charged fermions obtaining their masses via
the Higgs mechanism. The $0\nu \beta \beta $ decay amplitude is
proportional to the effective Majorana mass $\left\vert m_{\beta \beta
}\right\vert $ defined as $\left\vert m_{\beta \beta }\right\vert
=\left\vert \sum_{i}U_{ei}^{2}m_{i}\right\vert $, and may be expressed in
terms of our model parameters and the parameters of the $\mathcal{U}_{\nu }$%
\ mixing matrix%
\begin{equation}
	\left\vert m_{\beta \beta }\right\vert =\left\vert \frac{2m_{1}}{3}\cos
	^{2}\theta +\frac{1}{3}\sqrt{m_{1}^{2}+\Delta m_{21}^{2}}e^{\frac{i}{2}%
		\alpha _{21}}+\frac{2}{3}\sin ^{2}\theta \sqrt{m_{1}^{2}+\Delta m_{31}^{2}}%
	e^{\frac{i}{2}(\alpha _{31}-2\sigma )}\right\vert
\end{equation}%
Notice that the relevance of the absolute mass scale in $0\nu \beta \beta $%
\ experiments arise from the dependence of $\left\vert m_{\beta \beta
}\right\vert $ on $m_{i}$.
\begin{figure}[h]
	\begin{center}
		\includegraphics[width=.52\textwidth]{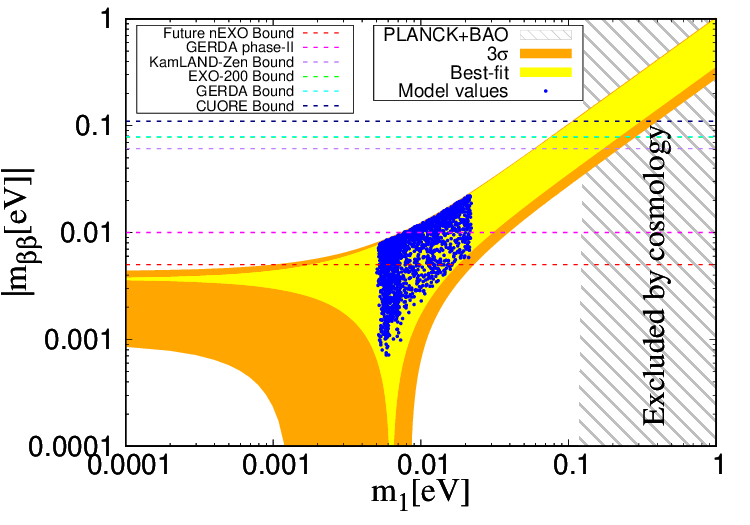}
	\end{center} \vspace{-2em} \caption{$\left\vert m_{\protect\beta
			\protect\beta }\right\vert $ as a function of $m_{1}$. The vertical
		gray dashed region indicates the upper limit on the sum of the three
		light neutrino masses from Planck+BAO data.} \label{f4}
\end{figure}
Although $0\nu \beta \beta $ decay has not been observed, there are
dozens of running and forthcoming experiments around the world
setting as
their objective the detection of this process. The current limits on $%
\left\vert m_{\beta \beta }\right\vert $ come from the KamLAND-Zen \cite{C7}%
, CUORE \cite{C8}, GERDA \cite{C9} and EXO \cite{C10} experiments
corresponding to $\left\vert m_{\beta \beta }\right\vert <(0.061-0.165)%
\mathrm{eV}$, $\left\vert m_{\beta \beta }\right\vert <(0.075-0.35)\mathrm{eV%
}$, $\left\vert m_{\beta \beta }\right\vert <(0.079-0.180)\mathrm{eV}$ and $%
\left\vert m_{\beta \beta }\right\vert <(0.078-0.239)\mathrm{eV}$ respectively. Figure (\ref{f4}) shows the correlation between
$\left\vert m_{\beta \beta }\right\vert $ and the lightest neutrino
mass $m_{1}$ for normal mass hierarchy. This plot is obtained by
varying the oscillation parameters in their $3\sigma $ range while
the Majorana phases are varied in the range $[0\rightarrow 2\pi ]$.
The horizontal dashed lines represent the limits on $\left\vert
m_{\beta \beta }\right\vert $ from current $0\nu \beta \beta $ decay
experiments while the vertical gray region is disfavored by the
Planck+BAO data. From this figure, we extract our range of the
effective
Majorana mass%
\begin{equation}
	0.000715\lesssim \left\vert m_{\beta \beta }\right\vert \left( \mathrm{eV}%
	\right) \lesssim 0.022028  \label{mee}
\end{equation}%
The predictions for $\left\vert m_{\beta \beta }\right\vert $ are
far from the current sensitivities mentioned above, on the other
hand, the next-generation experiments such as GERDA Phase II, CUPID,
nEXO and SNO+-II will cover the values of $\left\vert m_{\beta \beta
}\right\vert $ in eq. (\ref{mee}) as they aim for sensitivities
around $\left\vert m_{\beta
	\beta }\right\vert \sim \left( 0.01-0.02\right) \mathrm{eV}$ \cite{C11}, $%
\left\vert m_{\beta \beta }\right\vert \sim \left( 0.006-0.017\right)
\mathrm{eV}$\ \cite{C12}, $\left\vert m_{\beta \beta }\right\vert \sim
\left( 0.008-0.022\right) \mathrm{eV}$ \cite{C13} and $\left\vert m_{\beta
	\beta }\right\vert \sim \left( 0.02-0.07\right) \mathrm{eV}$ \cite{C14}
respectively.
\section{Leptogenesis}
\label{sec6}
In this section, we investigate the generation of the baryon asymmetry of the
universe within our SUSY $SU(5)\times D_{4}\times U(1)$ model in the case of normal mass hierarchy. In this scenario, the presence of three RH neutrinos as the key ingredients for
small neutrino masses can also produce the BAU through the leptogenesis
mechanism. In this case, a lepton asymmetry $Y_{L}$\ (equally $B-L$
asymmetry $Y_{B-L}$) is generated through the out-of-equilibrium $CP$
violating decays of RH neutrinos $N_{i}^{c}$ (and their supersymmetric
partners in SUSY models) in the early universe. This lepton asymmetry is
then partially converted into the baryon asymmetry of the universe $Y_{B}$
via ($B+L$) violating sphaleron transitions \cite{R19}.

\begin{figure}[h]
	\begin{center}
		\includegraphics[width=.55\textwidth]{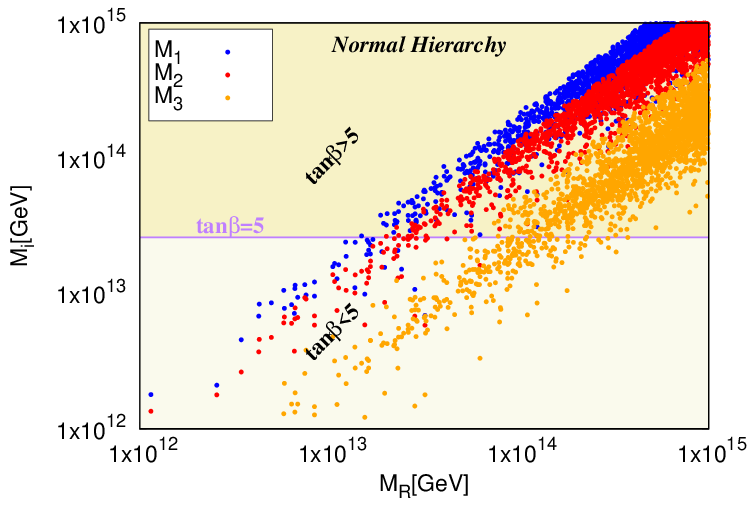}
	\end{center}
	\par
	\vspace{-2em} \caption{The RH neutrino masses ($M_{1}$, $M_{2}$ and
		$M_{3}$) as a function of the Majorana mass scale $M_{R}$.}
	\label{8}
\end{figure}

The excess of baryons over anti-baryons is evaluated through the baryon
asymmetry $Y_{B}$ relative to the entropy density $s$ or the baryon
asymmetry $\eta _{B}$ relative to the density of photons $n_{\gamma }$,
defined respectively as%
\begin{equation}
	Y_{B}=\frac{n_{B}-n_{\overline{B}}}{s}\quad \quad ,\quad \quad \eta _{B}=%
	\frac{n_{B}-n_{\overline{B}}}{n_{\gamma }}  \label{L1}
\end{equation}%
where $n_{B}$ and $n_{\overline{B}}$ are the number densities of baryons
and anti-baryons. The experimental values of these parameters obtained from the
latest data from the Planck satellite are given by $Y_{B}=(8.72\pm 0.08)\times 10^{-11}$ and $\eta _{B}=\left( 6.13\pm
0.04\right) \times 10^{-10}$ \cite{A23}. In order to perform an approximate
estimation of $Y_{B}$, we use the following two approaches:

\begin{itemize}
	\item It is well-known that when the right-handed
	neutrino mass spectrum is hierarchical, the contribution to the lepton
	asymmetry can be created only by the decay of the lightest RH neutrino \cite%
	{R13,C15,C16}. Since only the NH is allowed in our model, it is
	clear from eqs. (\ref{mas}) and (\ref{Mi}) as well as from figure (\ref{8}%
	) that $M_{3}$ is the lightest RH neutrino where the masses of $M_{1}$ and
	$M_{2}$ differ at most by a factor of 3 (with $%
	3M_{3}<M_{2},M_{1})$.
	
	\item Since all Majorana masses are above $%
	T=10^{12}(1+\tan ^{2}\beta )$ for $\tan \beta =5$ and $%
	\tan \beta =10$---used in the charged fermion sector to fit the
	experimental data---we perform our study in the one flavor
	approximation where all charged leptons are out-of-equilibrium and
	there is no difference between them at the time leptogenesis takes
	place.
\end{itemize}
Taking this two points into consideration, the magnitude of $B-L$
asymmetry generated by $N_{3}$ can be parameterized as follows
\cite{C17}
\begin{equation}
	Y_{B-L}=-\left( \varepsilon _{N_{3}}Y_{N_{3}}^{eq}+\varepsilon _{\tilde{N}%
		_{3}}Y_{\tilde{N}_{3}}^{eq}\right) \eta _{33}  \label{L2}
\end{equation}%
where $\varepsilon _{N_{3}}$ ($\varepsilon _{\tilde{N}_{3}}$) is the $CP$
asymmetry produced in the decay of $N_{3}^{c}$\ ($\tilde{N}_{3}^{c}$), $\eta
_{33}$ is the efficiency factor\footnote{%
	The efficiency factor in the general formula of the $B-L$ asymmetry is
	written as the sum over all the lepton flavors $\sum\limits_{\alpha ,\beta
	}\eta _{\alpha \beta }$. In the present study where we employ the one flavor
	approximation, $\eta _{33}$ is the only efficiency factor relevant for
	leptogenesis.} related to the washout of the $CP$ asymmetry $\varepsilon _{N_{3}}$ ($\varepsilon _{\tilde{N}_{3}}$) due to $N_{3}^{c}$ ($\tilde{N}_{3}^{c}$) decays, and $Y_{N_{3}}^{eq}$ is the number
density of $N_{3}$ over the entropy density ($n_{N_{3}}/s$) defined as \cite%
{C17}
\begin{equation}
	\left. Y_{N_{3}}^{eq}\right\vert _{T>>M_{3}}=\frac{135\zeta (3)}{%
		4\pi ^{4}g_{\ast }}
\end{equation}%
where $\zeta (3)$ denotes the Riemann zeta function and
$g_{\ast }$ is the number of spin-degrees of freedom in thermal equilibrium;
in MSSM $g_{\ast }=228.75$.

The first component to consider in our calculation is the source of the $CP$ asymmetry given by the $CP$ violating parameters $\varepsilon _{N_{i}}$ and
$\varepsilon _{\tilde{N}_{i}}$ in $N_{i}^{c}$ and $\tilde{N}_{i}^{c}$
decays, averaged over the different decay channels $N_{i}^{c}\rightarrow
LH_{u},$ $\tilde{L}\tilde{H}_{u}$ and $\tilde{N}_{i}^{c}\rightarrow \tilde{L}%
H_{u},$ $L\tilde{H}_{u}$ respectively. These RH neutrinos and their
superpartners decay, with decay rates that reads respectively as%
\begin{equation}
	\Gamma _{N}=\Gamma \left( N_{i}^{c}\rightarrow LH_{u}\right) +\Gamma \left(
	N_{i}^{c}\rightarrow \tilde{L}\tilde{H}_{u}\right) \text{ and }\Gamma _{%
		\tilde{N}}=\Gamma \left( \tilde{N}_{i}^{c}\rightarrow \tilde{L}H_{u}\right)
	+\Gamma \left( \tilde{N}_{i}^{c}\rightarrow L\tilde{H}_{u}\right)
\end{equation}%
while the $CP$ violating parameters are given by%
\begin{equation}
	\varepsilon _{N_{i}}=\frac{\Gamma _{N}-\bar{\Gamma}_{N}}{\Gamma _{N}+\bar{%
			\Gamma}_{N}}  \text{ and }  \varepsilon _{\tilde{N}_{i}}=\frac{\Gamma _{\tilde{N}%
		}-\bar{\Gamma}_{\tilde{N}}}{\Gamma _{\tilde{N}}+\bar{\Gamma}_{\tilde{N}}}
\end{equation}%
In SUSY models, the effects from the superparticles produce relatively small
corrections to the BAU \cite{C18}. Therefore, by ignoring supersymmetry
breaking\footnote{%
	For cases where SUSY can not be ignored see \cite{C19,C20}.}---as a result
of which the RH neutrinos and their superpartners have equal masses $%
M_{N_{i}}=M_{\tilde{N}_{i}}$, equal decay rates $\Gamma _{N_{i}}=\Gamma _{%
	\tilde{N}_{i}}$ and equal $CP$ asymmetries $\varepsilon _{N_{i}}=\varepsilon
_{\tilde{N}_{i}}$ \cite{C17}---we can factorize by the $CP$ asymmetry in eq.
(\ref{L2}) as $Y_{B-L}=-\varepsilon _{N_{3}}\left( Y_{N_{3}}^{eq}+Y_{\tilde{N%
	}_{3}}^{eq}\right) \eta _{33}$. Likewise, when the equilibrium densities
for leptons and sleptons are equal $Y_{N_{i}}^{eq}\approx Y_{\tilde{N}%
	_{i}}^{eq}$, we find that the $B-L$ asymmetry parameter $Y_{B-L}$ is
enhanced by a factor of $2$. Bringing together all these effects, the $CP$
asymmetry can be explicitly expressed in the one flavor approximation as%
\begin{equation}
	\varepsilon _{N_{3}}=\frac{1}{8\pi} \sum_{j=1,2}\frac{Im\left[
		\left( \mathcal{Y}_{\nu }\mathcal{Y}_{\nu }^{\dagger }\right) _{j3}^{2}%
		\right] }{\left( \mathcal{Y}_{\nu }\mathcal{Y}_{\nu }^{\dagger }\right) _{33}%
	}f\left( \frac{M_{j}}{M_{3}}\right)  \label{cpa}
\end{equation}%
where $f\left( x\right) =\sqrt{x}\left( 1-\left( 1+x\right) \ln \left[
\left( 1+x\right) /x\right] \right) $ and\emph{\ }$\mathcal{Y}_{\nu }$\emph{%
	\ }is the neutrino Yukawa coupling matrix in the basis where the Majorana
mass matrix $m_{M}$ and the Yukawa matrix of the charged leptons $\mathcal{Y}%
_{e}$ are both diagonal. However, as explained in the appendix
\ref{app2}, the contribution of the mixing matrix that diagonalizes
$\mathcal{Y}_{e}$ leads to $CP$ asymmetry of order $\left\vert
\varepsilon _{N_{3i}}\right\vert \sim
\mathcal{O}(10^{-12}-10^{-10})$ which suppress the value of the
baryon asymmetry $Y_{B}$. Therefore, in order to meet the
requirements of a successful leptogenesis that produces the
experimental values of $Y_{B}$, we add a correction to the leading
order Dirac Yukawa matrix in eq. (\ref{md}). To account for this
correction, we
introduce a new flavon field $\omega $ which transforms as $1_{+-}$ under $%
D_{4}$ with zero $U(1)$ charge, we have
\begin{equation}
	\delta W_{D}= \frac{\lambda _{9}}{\Lambda }N_{3,2}^{c}F_{2,3}H_{5}\omega  \label{D}
\end{equation}%
where $\lambda _{9}$ is a complex coupling constant $\lambda
_{9}=\left\vert \lambda _{9}\right\vert e^{i\phi _{\omega }}$. This
effective coupling is obtained from the following
renormalizable superpotential
\begin{equation}
	W_{D}^{ren}=N_{3,2}^{c}F_{2,3}X_{5}+\overline{X}_{5}H_{5}\omega
\end{equation}%
where $X_{5}$\ is a messenger field that transforms as $SU(5)$ quintet, $%
D_{4} $ singlet $1_{+-}$ and has a $U(1)$ charge equals to $-8$. The
contribution of $\delta W_{D}$ is small and will not provide
any considerable effect in the obtained neutrino masses and mixing. When the
flavon field $\omega $ acquires its VEV as $\left\langle \omega
\right\rangle =\upsilon _{\omega }$, we end up with the total Yukawa mass
matrix\footnote{%
	Notice that the total light neutrino mass matrix involving the small
	correction $\delta Y_{D}$ is almost similar to the one in eq. (\ref{wn}) and
	yields approximately to the same neutrino phenomenology.}
\begin{equation}
	\mathcal{Y}_{D}=Y_{D}+\delta Y_{D}=\frac{m_{D}}{\upsilon _{u}}+\delta
	Y_{D}=\left(
	\begin{array}{ccc}
		\lambda _{1} & 0 & 0 \\
		0 & \lambda _{1} & 0 \\
		0 & 0 & \lambda _{1}%
	\end{array}%
	\right) +\kappa e^{i\phi _{\omega }}\left(
	\begin{array}{ccc}
		0 & 0 & 0 \\
		0 & 0 & 1 \\
		0 & 1 & 0%
	\end{array}%
	\right)
\end{equation}%
where $\kappa =\frac{\left\vert \lambda _{9}\right\vert \upsilon _{\omega }%
}{\Lambda }$ is a free parameter which should be small ($\kappa <<1$) in
order to produce the correct BAU. Taking into account this correction, the
total Yukawa neutrino mass matrix is defined as $\mathcal{Y}_{\nu }=\mathcal{%
	U}_{\nu }^{\dagger }\mathcal{Y}_{D}$. Thus, after calculating the product $%
\mathcal{Y}_{\nu }\mathcal{Y}_{\nu }^{\dagger }$ in the basis where the
Majorana mass matrix is diagonal, the $CP$ asymmetry parameter $\varepsilon
_{N_{3}}$ corresponding to the lightest RH neutrino $N_{3}$ is given
approximately by
\begin{eqnarray}
	\varepsilon _{N_{3}} &\simeq&\frac{\kappa ^{2}}{9\pi }\cos ^{2}\phi _{\omega }%
	\left[ 2\sin ^{2}(2\theta )\sin ^{2}(\sigma -\frac{\alpha _{31}}{2})f\left(
	\frac{\tilde{m}_{1}}{\tilde{m}_{3}}\right) \right.   \notag \\
	&&+\left. \sin ^{2}\theta \sin ^{2}\left( \sigma +\frac{(\alpha _{21}-\alpha
		_{31})}{2}\right) f\left( \frac{\tilde{m}_{2}}{\tilde{m}_{3}}\right) \right]
	\label{eps}
\end{eqnarray}%
where $\tilde{m}_{i}$ are the washout mass parameters expressed as $\tilde{m}%
_{i}=\upsilon _{u}^{2}\frac{\left( \mathcal{Y}_{\nu
	}\mathcal{Y}_{\nu }^{\dagger }\right) _{ii}}{M_{i}}$.

\begin{figure}[h]
	\begin{center}
		\includegraphics[width=.44\textwidth]{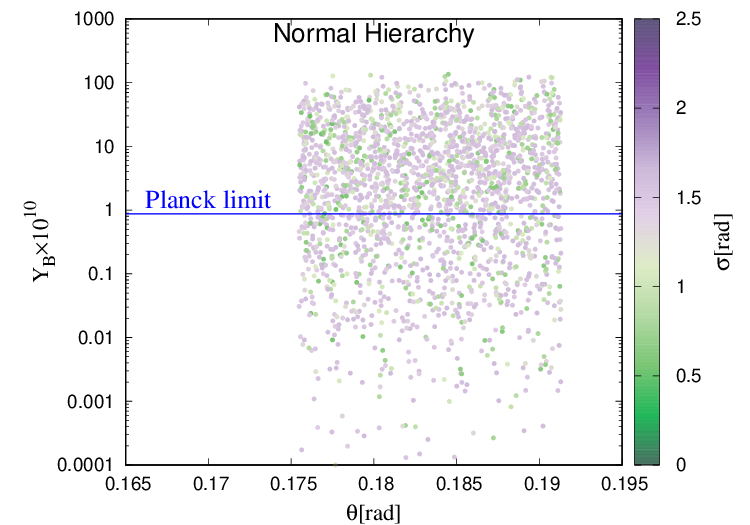}\quad %
		\includegraphics[width=.44\textwidth]{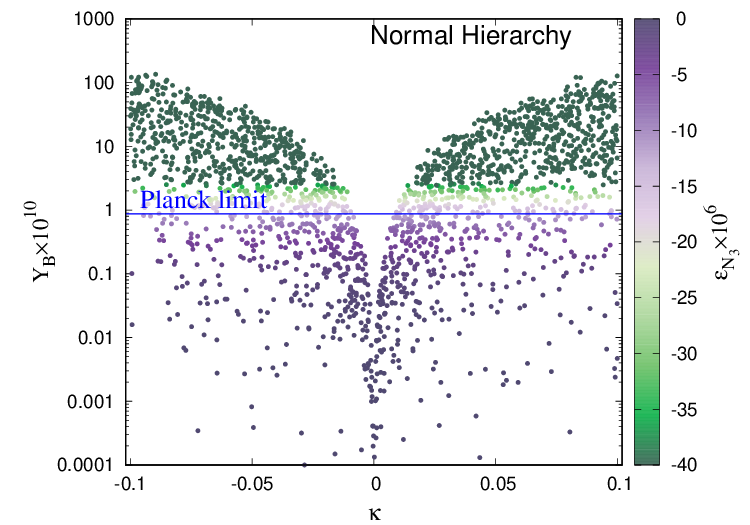}
	\end{center}
	\par
	\vspace{-2em}
	\caption{Left panel: $Y_{B}$ as a function of the
		trimaximal parameters $\protect\theta $ and $\protect\sigma $. Right panel: $Y_{B}$ as a function of the parameter $\protect\kappa$ with the color code indicating the allowed range of the $CP$ asymmetry parameter $\protect%
		\varepsilon _{N_{3}}$. The horizontal blue
		line corresponds to the Planck bound on $Y_{B}$.}
	\label{9}
\end{figure}

\begin{figure}
	\centering
	\includegraphics[width=.44\textwidth]{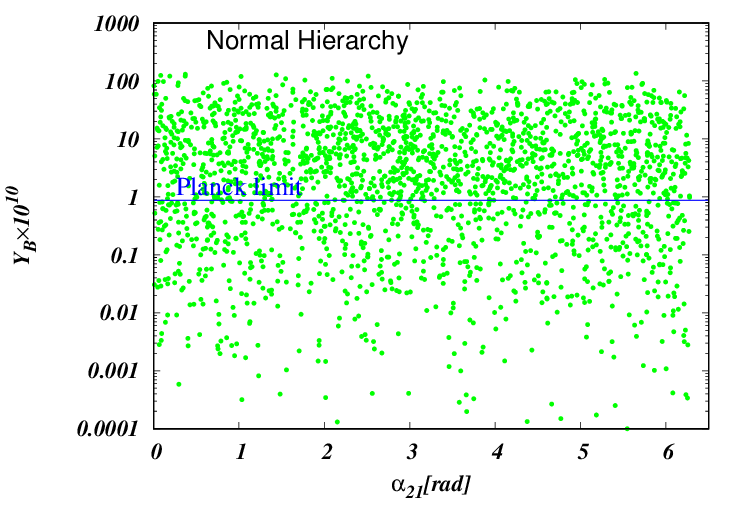}
	\includegraphics[width=.44\textwidth]{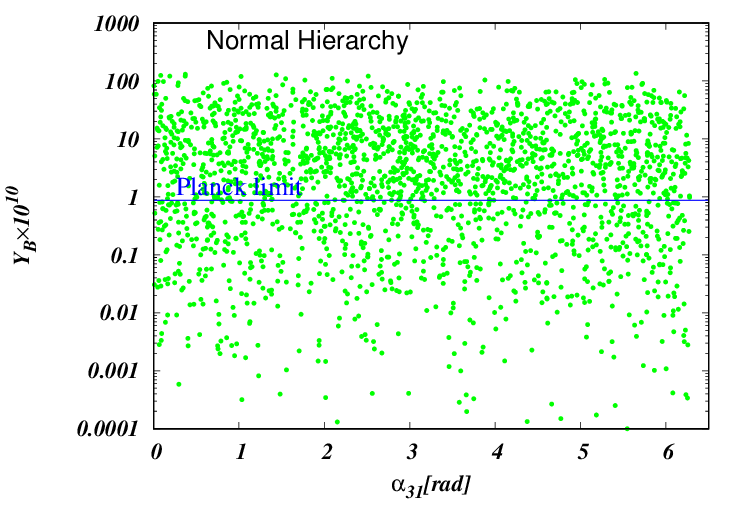}
	\includegraphics[width=.44\textwidth]{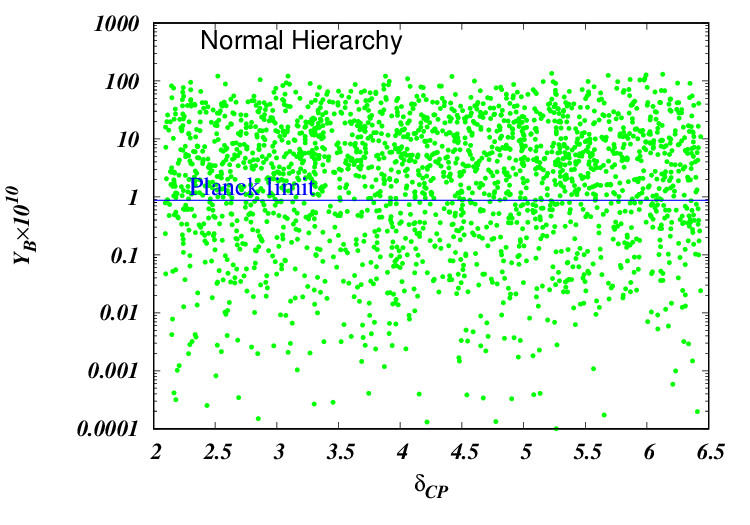}
	\includegraphics[width=.44\textwidth]{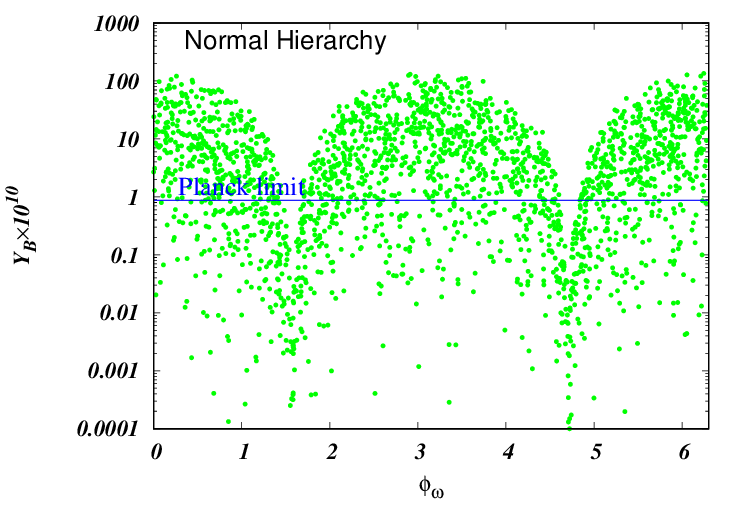}
	\vspace{-2em}
	\caption{The baryon asymmetry $Y_{B}$ as a function of Majorana phases $%
		\protect\alpha _{21}$ (top left), $\protect\alpha _{31}$ (top right), Dirac
		CP phase $\protect\delta _{CP}$ (bottom left) and the high energy CP phase $%
		\protect\phi _{\protect\omega }$ (bottom right). The horizontal blue line
		corresponds to the Planck bound.}
	\label{10}
\end{figure}

\begin{figure}[h]
	\centering
	\centering\includegraphics[width=.44\textwidth]{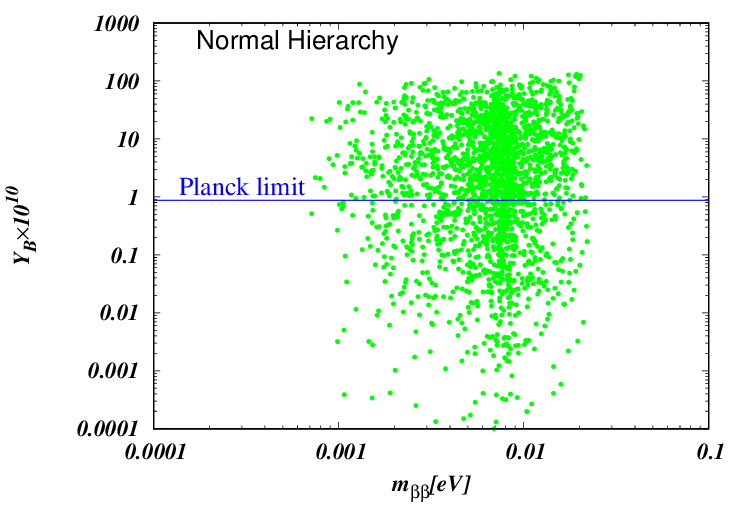}%
	\includegraphics[width=.44\textwidth]{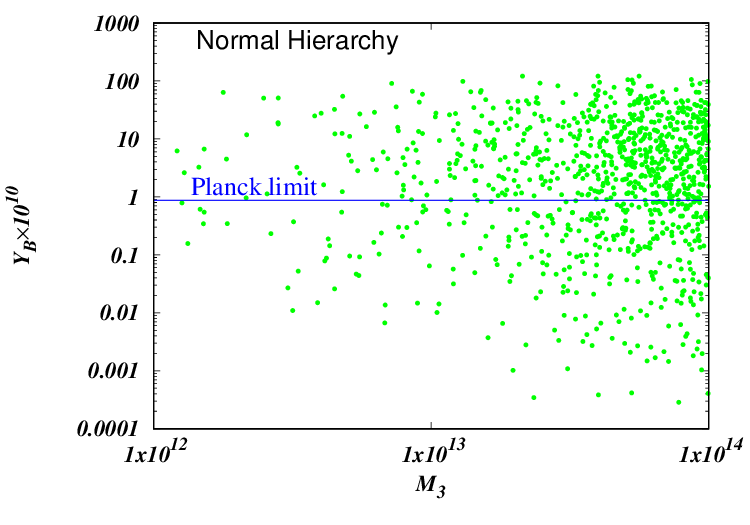} \includegraphics[width=.44%
	\textwidth]{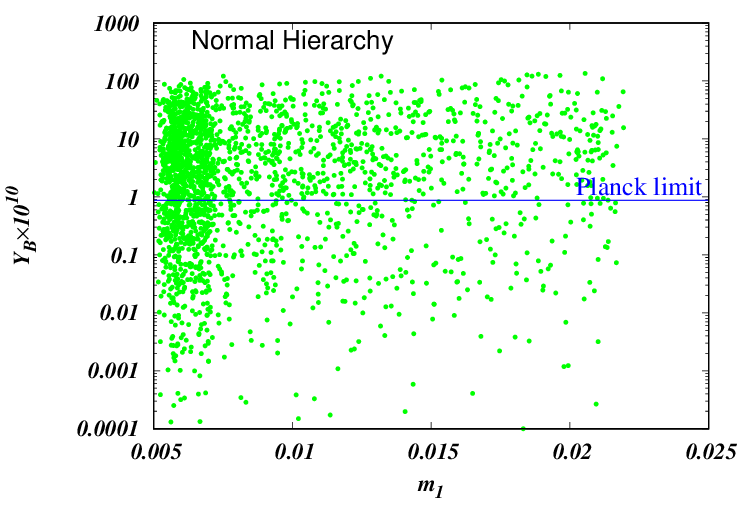}
	\vspace{-2em} \caption{The baryon asymmetry $Y_{B}$ as a function
		of the effective Majorana mass $m_{\protect\beta \protect\beta }$
		(top left), the lightest RH neutrino mass $M_{3}$ (top right) and
		the lightest neutrino mass $m_{1}$\ (bottom). The horizontal blue
		line corresponds to the Planck bound.} \label{12}
\end{figure}

The second component to address in this computation is the
efficiency factor $\eta _{33}$. A good approximation is to consider
the region of RH neutrino masses smaller than $10^{14}$
$\mathrm{GeV}$, preventing possible washout effects from $\Delta
L=2$ scattering processes. In this case, the efficiency factor $\eta
_{33}$ can be expressed approximately as a function
of the washout mass parameter $\tilde{m}_{3}$ as \cite{C17}%
\begin{equation}
	\eta _{33}\approx \left( \frac{3.3\times 10^{-3}\text{\textrm{eV}}}{\tilde{m}%
		_{3}}+\left( \frac{\tilde{m}_{3}}{0.55\times 10^{-3}\text{\textrm{eV}}}%
	\right) ^{1.16}\right) ^{-1}\quad
\end{equation}
Notice here that the smallness of the parameter $\kappa
<<\lambda _{1}$ implies that the washout mass parameters become
approximately
identical $\tilde{m}_{i}\approx m_{i}$ and hence $\tilde{m}_{3}\approx m_{3}$%
. Moreover, since the neutrino mass $m_{3}$ has values close to $%
0.5\times 10^{-1}$ \textrm{eV} as given in eq. (\ref{m1}), then the
efficiency factor $\eta _{33}$ in our model is roughly $\eta
_{33}\approx 0.5\times 10^{-2}$.\newline
Let us now derive the expression of the baryon asymmetry parameter
$Y_{B}$. This parameter is related to lepton asymmetry $Y_{B-L}$ given in eq. (\ref%
{L2}) through sphaleron transitions, we have \cite{C21}%
\begin{equation}
	Y_{B}\approx \left( \frac{8N_{f}+4N_{H}}{22N_{f}+13N_{H}}\right)
	Y_{B-L}=-2\left( \frac{8N_{f}+4N_{H}}{22N_{f}+13N_{H}}\right)
	Y_{N_{3}}^{eq}\varepsilon _{N_{3}}\eta _{33}
\end{equation}%
where $N_{f}=3$ is the number of fermion generations and $N_{H}=2$ is the
number of Higgs doublets in the MSSM. Accordingly, the amount of the
baryon asymmetry generated in the present model is given by
\begin{equation}
	Y_{B}\approx -1.266\times 10^{-3}\varepsilon _{N_{3}}\eta _{33}
\end{equation}%
Therefore, $Y_{B}$ in our model depends on the
trimaximal parameters $\theta $ and $\sigma $, the light neutrino masses $%
m_{i}$, the Majorana phases $\alpha _{31}$ and $\alpha _{21}$, as well as $%
\kappa $ and the phase $\phi _{\omega }$ coming from the extra
contribution in the Dirac mass matrix. Using the ranges of the
parameters $\theta $ and $\sigma $ restricted by the neutrino oscillation
data, we show in the left panel of figure (\ref{9}) the correlation between $Y_{B}$, $\theta $ and $%
\sigma $. We observe that there are many scattered points that correlate $%
\theta $ and $\sigma $ with the Planck bound on $Y_{B}$.
On the other hand, since the parameters $\kappa $ and $\phi _{\omega }$
are not controlled by the neutrino oscillation data, we allow them
to vary in the ranges $\left[ -0.1\rightarrow0.1\right] $ and $\left[ 0\rightarrow2\pi
\right] $ respectively. Then, we plot in the right panel of figure (\ref{9}) the correlation between $Y_{B}$
and $\kappa $ where the color palette corresponds to the absolute
value of the $CP$ asymmetry parameter $\varepsilon _{N_{3}}$. We
find that the observed baryon
asymmetry correspond to $\kappa $ in the range $\left[ -0.1\rightarrow-0.0085%
\right] \cup \left[ 0.009\rightarrow0.1\right] $ and $\left\vert \varepsilon _{N_{3}}\right\vert $ in the range $[1.302\rightarrow1.433]\times 10^{-5}$.
\newline Furthermore, it is clear from the $CP$ asymmetry parameter
$\varepsilon _{N_{3}}$ in eq. (\ref{eps}) that the source of $CP$
violation in the lepton sector could arise from the interplay
between the low energy $CP$ phases
(Dirac and Majorana phases $\delta _{CP}$, $\alpha _{31}$ and $\alpha _{21}$%
) and the high energy $CP$ phase $\phi _{\omega }$ originated from
the complex coupling constant $\lambda _{9}$ in the Dirac mass matrix; see
eq. (\ref{D}). Therefore, we plot in figure (\ref{10}), the baryon asymmetry parameter $%
Y_{B}$ as a function of the low energy $CP$ phases ($\alpha _{31}$,
$\alpha _{21}$ and $\delta _{CP}$) and the high energy $CP$ phase
$\phi _{\omega }$ which is the key ingredient for generating the observed range of $Y_{B}$.
We observe that the ranges of the Majorana phases (top panels) and the Dirac phase (bottom left panel) are not constrained compared to
their inserted intervals, nevertheless, the scattered points---including the $CP$ conserving values of the Majorana phases $\alpha _{31},\alpha
_{21}=0,\pi $---are
consistent with the Planck limit on $Y_{B}$. However, even in the
case of these $CP$ conserving values, $CP$ violation is guaranteed by the high energy $CP$ phase $\phi
_{\omega }$. For this reason, we plot in the bottom right panel of figure (\ref{10}) the correlation between $Y_{B}$ and $\phi
_{\omega }$ where we find that $\phi _{\omega }$ vary within the range $%
0\lesssim \phi _{\omega }\lesssim 6.279$, while the $CP$
conserving values $\phi _{\omega }=\frac{\pi }{2}$ and $\phi
_{\omega }=\frac{3\pi }{2}$ as well as the regions around them are
excluded (the sections of the blue line without any points). Therefore, this source of $CP$ violation plays a crucial
role in generating the baryon asymmetry in the present model.
\newline From another point of view, since both $Y_{B}$ and
$m_{\beta \beta }$ depend on the Majorana phases $\alpha _{31}$ and
$\alpha _{21}$, there exists a correlation between the effective
Majorana mass $m_{\beta \beta }$ which governs the $0\nu \beta \beta
$ process and the baryon asymmetry parameter $Y_{B}$. Therefore, we
display in the top left of figure (\ref{12}) $Y_{B}$ as a function of
$m_{\beta \beta }$ where we observe that there are several points
satisfying the Planck limit on the baryon asymmetry parameter.
Besides $m_{\protect\beta \protect\beta }$, the correlation of $Y_{B}$ with the lightest RH neutrino mass $M_{3}$ and the lightest neutrino mass $m_{1}$ is displayed, respectively, in the top right panel and the bottom panel of figure (\ref{12}), where we observe that there are several allowed points for both $M_{3}$ and $m_{1}$ within the Planck limit on $Y_{B}$.
\section{Summary and conclusion}
\label{sec7}
In this work, we have presented a model with a $D_{4}$ family
symmetry to
explain the fermion flavor structures in the framework of supersymmetric $%
SU(5)$ grand unified theory. Besides the $SU(5)\times D_{4}$ model proposed in ref. \cite{D1} -- which was merely an implementation of the $D_{4}$ in $SU(5)$ -- this is the first comprehensive study of a
four-dimensional $SU(5)$ GUT with a flavor symmetry that does not
include triplet irreducible representations. To establish a thorough analysis of this model, we have enlarged the field
content of the usual scalar and matter sectors of SUSY $SU(5)$ GUT.
Explicitly, we have added three RH neutrinos to generate neutrino
masses via the type I seesaw mechanism, heavy messenger fields to
make the model renormalizable at the GUT scale, higher dimensional
Higgs multiplets to produce realistic quark-lepton Yukawa coupling
ratios, and gauge singlet flavon fields to give rise to the observed
fermion mass spectrum and mixing through the spontaneous symmetry
breaking of the flavor group. Moreover, after adding these fields,
an additional $U(1)$ symmetry is imposed to control the invariance of
the superpotentials in the quark and lepton sectors, and also the dangerous $d=4$ and $d=5$ proton decay
operators.\newline
\newline
Integrating out the heavy messenger fields from the renormalizable
superpotentials gives rise to higher-dimensional effective operators
responsible for the fermion flavor structures. Moreover, to go
beyond the minimal $SU(5)$ relation
$\mathcal{Y}_{e}^{T}=\mathcal{Y}_{d}$ as well as the popular GJ
relations which are disfavored by the experimental results, we have
considered the CG factors $y_{e}/y_{d}=4/9$ and $y_{\mu }/y_{s}=9/2$
which are realized through the coupling of messenger fields with
higher 24- and 45-dimensional Higgs fields and the flavon fields.
This has led to the double ratio $\frac{y_{\mu
}}{y_{s}}\frac{y_{d}}{y_{e}}\simeq 10.12$ which is in good
agreement with the phenomenological value at GUT scale. We have
performed a numerical analysis in the down and charged lepton Yukawa
sector where we have fixed our model parameters -- the free
parameters in the entries of the Yukawa matrices -- and provided an
accurate fit to the mixing angles, the Yukawa couplings and the
Dirac $CP$ phase of the quark sector at the GUT scale.\newline
\newline
The small neutrino masses are generated via the type I seesaw
mechanism where the Dirac and Majorana mass matrices arise from
renormalizable terms. The resulting neutrino mass matrix is of the
trimaximal mixing form which is compatible with current neutrino
data. By using the $3\sigma $ experimental range of $\sin ^{2}\theta
_{13}$ for both neutrino mass hierarchies we derived the range of
the trimaximal mixing parameter $\theta $ where we found that only
the normal mass hierarchy is allowed. Therefore, we have carried out
our numerical study in this regime where we found that our model
allows for $\theta _{13}\neq 0$ and $\theta _{23}<\pi /4$ as well as
excludes the conserving values of the Dirac neutrino $CP$ phase
$\delta _{CP} $. We have explored the neutrino parameter space and
showed numerically the predicted ranges of the non-oscillatory
observables $m_{\beta }$, $m_{\beta \beta }$ and $\Sigma m_{i}$
that fit the $3\sigma $ experimental range of the mixing angles and
the mass squared splittings. In particular, the predicted values of
$m_{\beta \beta }$ are testable at future neutrinoless double beta
decay experiments.\newline \newline Since the low energy $CP$
violation which manifest itself in the mixing matrix in the form of
the Dirac and Majorana phases is not sufficient to describe the BAU,
we have added an extra effective operator in the neutrino sector to
produce the observed BAU via the leptogenesis mechanism. This
operator which involves a new flavon field $\omega $ is obtained, as
in the quark sector, by integrating out heavy messenger fields. Its
contribution serves as a correction that perturbs the structure of
the Dirac mass matrix while the high energy $CP$ phase $\phi
_{\omega }$ that arises from the complex coupling constant in this
operator is a new source of $CP$ violation. Therefore, we have
performed a numerical study to estimate the values of the $CP$
asymmetry parameter $\varepsilon _{N_{3i}}$ that are consistent with
the baryon asymmetry parameter $Y_{B}$. We have focused on the
unflavored leptogenesis approximation scenario under which the range of the lightest RH neutrino mass is given by $M_{3}\left(
\mathrm{GeV}\right)
\in \left[ 2.6\times 10^{13}\rightarrow 10^{14}\right] $. We found that the $%
CP$\ asymmetry parameter $\varepsilon _{N_{3}}$ is mainly related to
the high energy $CP$ phase $\phi _{\omega }$. Therefore, we showed
through scatter plots that the $CP$ conserving values $\phi
_{\omega }=\frac{\pi }{2}$ and $\phi _{\omega }=\frac{3\pi }{2}$
as well as the regions around them are excluded, while
the lepton asymmetry parameter $\varepsilon _{N_{3i}}$ must be of order $%
\left\vert \varepsilon _{N_{3i}}\right\vert \sim O(10^{-5})$ to
satisfy the Planck limit on $Y_{B}$.
\appendix
\section{Messenger sector}
\label{app1}
\begin{figure}[h]
	\centering
	\includegraphics[width=.26\textwidth]{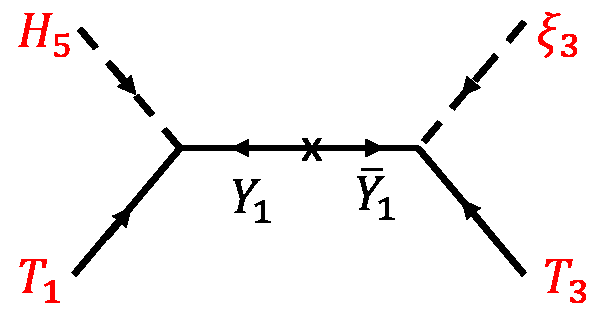} \includegraphics[width=.26%
	\textwidth]{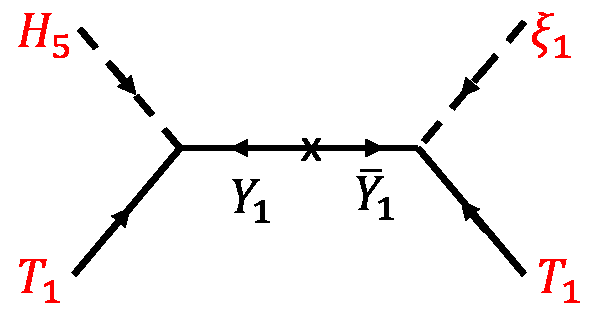} \includegraphics[width=.26\textwidth]{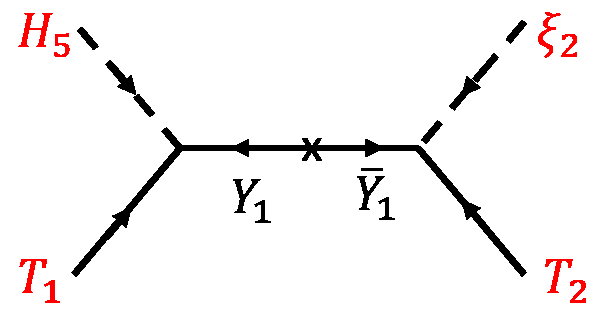} %
	\includegraphics[width=.26\textwidth]{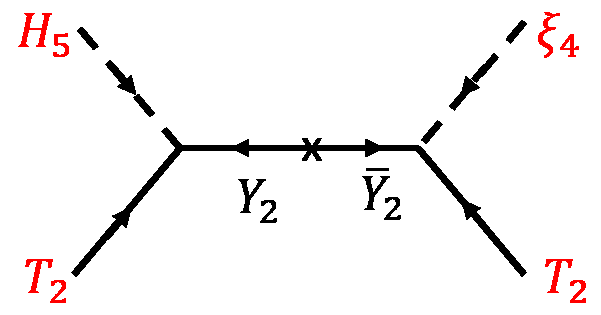} \includegraphics[width=.26%
	\textwidth]{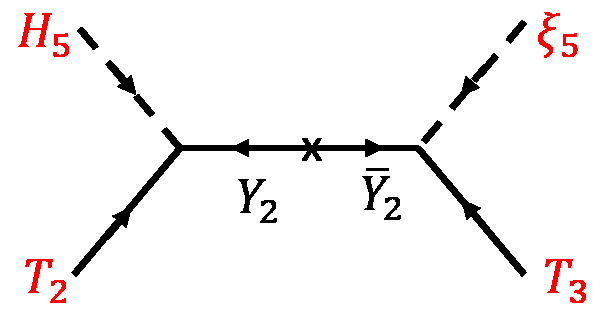} \includegraphics[width=.22\textwidth]{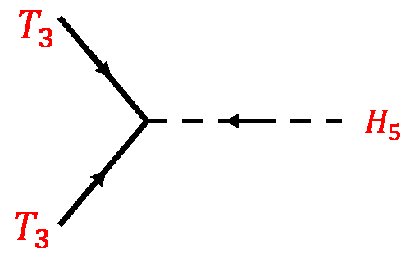}
	\caption{Diagrams inducing effective operators for the
		up-type quark sector. The last diagram implies
		that the top quark mass originate from the renormalizable operator $%
		T_{3}T_{3}H_{5}.$} \label{d1}
\end{figure}

\begin{figure}[H]
	\centering
	\includegraphics[width=.30\textwidth]{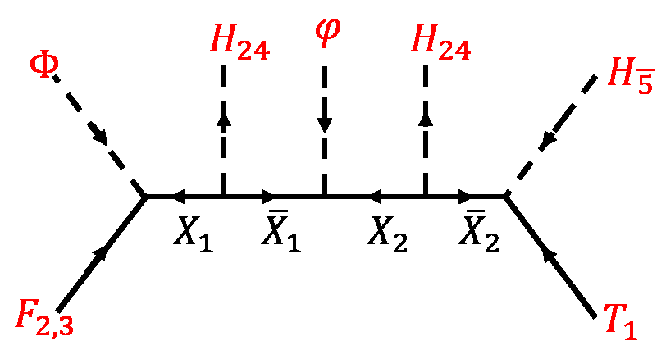} \includegraphics[width=.23%
	\textwidth]{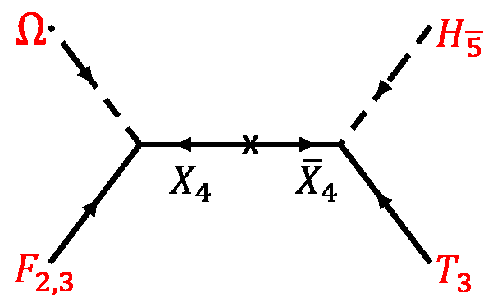} \includegraphics[width=.30\textwidth]{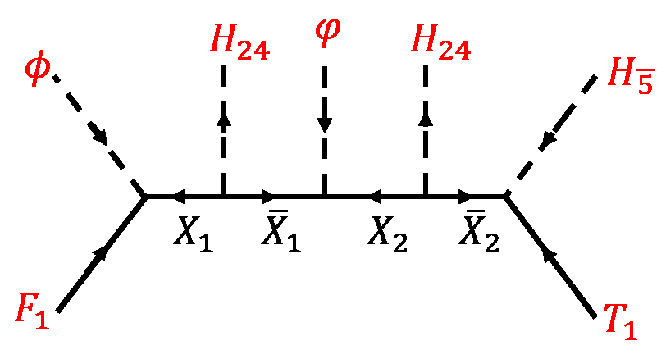} %
	\includegraphics[width=.30\textwidth]{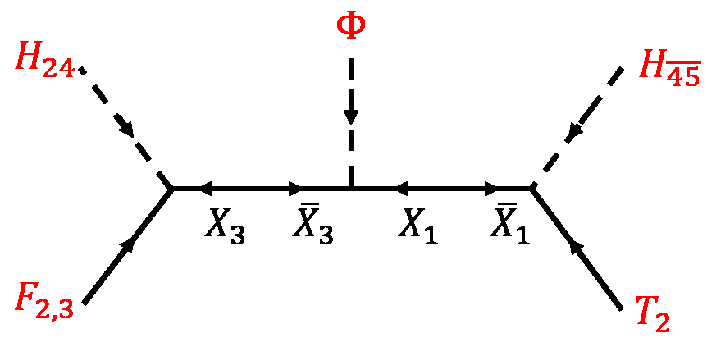} \includegraphics[width=.245%
	\textwidth]{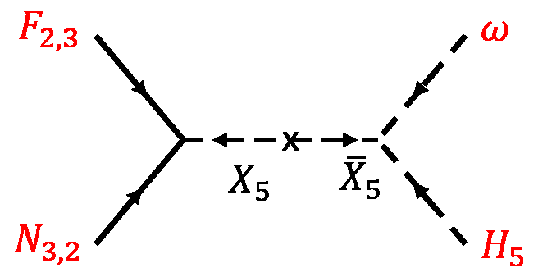}
	\caption{Diagrams inducing effective operators
		for the down-type quark and lepton sector. The last diagram is
		associated to the small Dirac correction.} \label{d2}
\end{figure}

In this Appendix we discuss the renormalizable superpotentials of
all the fermions including their Feynman diagrams to obtain the
higher dimensional operators relevant for the Yukawa mass matrices.
The complete list of the messenger field content including their
$SU(5)$ and $D_{4}$ representations as well as their
$U\mathbb{(}1\mathbb{)}$ charges is given in table (\ref{mes}).
To be precise, the messenger fields $Y_{i}$ are relevant for the up
quark sector while $X_{i}$ are involved in the down quark, the
charged lepton and the Dirac neutrino sectors.
\begin{table}[h]
	\centering$%
	\begin{tabular}{l||l|l|l|l|l|l|l}
		\hline
		Messenger fields & $X_{1}$ & $X_{2}$ & $\ \ \ \ X_{3}$ & $X_{4}$ & $X_{5}$ &
		$Y_{1}$ & $Y_{2}$ \\ \hline
		$SU(5)$ & $5$ & $5$ & $\ \ \ \ \ 5$ & $5$ & $5$ & $10$ & $10$ \\ \hline
		$D_{4}$ & $1_{+,-}$ & $1_{+,+}$ & $\left(
		\begin{array}{c}
			0 \\
			X_{3}%
		\end{array}%
		\right) $ & $1_{+,-}$ & $1_{+,-}$ & $1_{+,-}$ & $1_{+,-}$ \\ \hline
		$U(1)$ & $-4$ & $10$ & $-13$ & $8$ & $-8$ & $2$ & $-4 $ \\ \hline
	\end{tabular}
	$
	\caption{Messenger fields relevant in our model and their $D_{4}$
		representations as well as their $U(1)$ charges. The messenger fields $X_{i}$
		live in $5$-dimensional representation while $Y_{i}$ live in $10$%
		-dimensional representation. We assume that their mass is around the GUT
		scale.}
	\label{mes}
\end{table}
The renormalizable superpotential invariant under $D_{4}\times U(1)$
associated to the up quarks reads as
\begin{eqnarray}
	W_{up}^{Ren} &=&H_{5}T_{1}Y_{1}+\overline{Y}_{1}T_{1}\xi
	_{1}+H_{5}T_{1}Y_{1}+\overline{Y}_{1}T_{2}\xi _{2}+H_{5}T_{1}Y_{1}  \notag \\
	&&+\overline{Y}_{1}T_{3}\xi _{3}+H_{5}T_{2}Y_{2}+\overline{Y}_{2}T_{2}\xi
	_{4}+H_{5}T_{2}Y_{2}+\overline{Y}_{2}T_{3}\xi _{5}
\end{eqnarray}
where we have omitted the coupling constants from all terms for simplicity.
The couplings in this superpotential are illustrated by the Feynman diagrams
provided in figure (\ref{d1}). After integrating out the messenger fields $%
Y_{i}$ and $\bar{Y}_{i}$ from $W_{up}^{Ren}$ we obtain the
effective superpotential responsible for the masses of the up
quarks given in eq. (\ref{wu}).
As for the down-type quark and charged lepton sector, the renormalizable
superpotential involving the five-plets messenger fields $X_{i}$ and $%
\overline{X}_{i}$ is given by%
\begin{eqnarray}
	W_{d,e}^{Ren} &=&F_{1}\phi X_{1}+X_{1}H_{24}\overline{X}_{1}+\overline{X}%
	_{1}\varphi X_{2}+X_{2}H_{24}\overline{X}_{2}+\overline{X}_{2}H_{\overline{5}%
	}T_{1}+F_{2,3}\Phi X_{1}  \notag \\
	&&+X_{1}H_{24}\overline{X}_{1}+\overline{X}_{1}\varphi X_{2}+X_{2}H_{24}%
	\overline{X}_{2}+\overline{X}_{2}H_{\overline{5}}T_{1}+F_{2,3}H_{24}X_{3}
	\label{Wde} \\
	&&+\overline{X}_{3}\Phi X_{1}+\overline{X}_{1}H_{\overline{45}%
	}T_{2}+F_{2,3}\Omega X_{4}+\overline{X}_{4}H_{\overline{5}}T_{3}  \notag
\end{eqnarray}%
while the renormalizable terms relevant for the effective operator
responsible for generating a successful BAU is given as follows%
\begin{equation}
	W_{D}^{ren}=N_{3,2}^{c}F_{2,3}X_{5}+\overline{X}_{5}H_{5}\omega   \label{Wdn}
\end{equation}
Once more, integrating out the these heavy messenger fields give
rise to the effective superpotentials of the down quarks, the
charged leptons and the Dirac neutrino; see eqs. (\ref{wd}) and
(\ref{D}). The mass terms of the
messenger fields takes the form $W_{MF}=M_{X_{i}}X_{i}\overline{X}%
_{i}+M_{Y_{j}}Y_{j}\overline{Y}_{j}$ where $i=1,2,..5$ and $j=1,2$.
In fact, $\overline{X}_{i}$ and $\overline{Y}_{j}$ are the
corresponding fields of $X_{i}$ and $Y_{j}$, they are hosted by the
$SU(5)$ representations $\overline{5}$ and $10$ respectively, and
they have the same $D_{4}$ representations as their partners but
with opposite $U(1)$ charges. The Feynman diagrams relevant for the
superpotetials (\ref{Wde}) and (\ref{Wdn}) are illustrated in figure
(\ref{d2}).
\section{$CP$ asymmetry from the charged lepton mixing and $d=6$ Dirac
	operators} \label{app2}
In this appendix, we show that the contribution of the charged
lepton mixing matrix to the Yukawa mass matrix in eq. (\ref{md}),
before adding the correction $\delta Y_{D}$, cannot accommodate the
observed value of the BAU. Starting with the $CP$ asymmetry formula
in eq. (\ref{cpa}) which can be explicitly expressed in the one flavor approximation as%
\begin{equation}
	\varepsilon _{N_{3}^{c}}=\frac{1}{8\pi }\left\{ \frac{Im\left[ \left(
		\mathcal{Y}_{\nu }\mathcal{Y}_{\nu }^{\dagger }\right) _{13}^{2}\right] }{%
		\left( \mathcal{Y}_{\nu }\mathcal{Y}_{\nu }^{\dagger }\right) _{33}}f\left(
	\frac{M_{1}}{M_{3}}\right) +\frac{Im\left[ \left( \mathcal{Y}_{\nu }%
		\mathcal{Y}_{\nu }^{\dagger }\right) _{23}^{2}\right] }{\left( \mathcal{Y}%
		_{\nu }\mathcal{Y}_{\nu }^{\dagger }\right) _{33}}f\left( \frac{M_{2}}{M_{3}}%
	\right) \right\},  \label{cpa2}
\end{equation}%
where $\mathcal{Y}_{\nu }=\mathcal{U}_{TM_{2}}^{\dagger }Y_{D}\mathcal{U}%
_{l} $ is the neutrino Yukawa coupling matrix in the basis where the
Majorana mass matrix $m_{M}$ and the Yukawa matrix of the charged leptons $%
\mathcal{Y}_{e}$\ are both diagonal---see section \ref{sec4}---with\footnote{%
	The diagonalization of $\mathcal{Y}_{e}$ given in eq. (\ref{ye}) is
	obtained as a function of $b_{11}$, $b_{12}$ and $b_{22}$. Then, we
	replace these
	parameters by their numerical values given in table (\ref{t6}) to produce $%
	\mathcal{U}_{l}$.}

\begin{equation}
	Y_{D}=\frac{m_{D}}{\upsilon _{u}}=\lambda _{1}\left(
	\begin{array}{ccc}
		1 & 0 & 0 \\
		0 & 1 & 0 \\
		0 & 0 & 1%
	\end{array}%
	\right) \quad ,\quad \mathcal{U}_{l}=\left(
	\begin{array}{ccc}
		\frac{8b_{11}-81b_{22}}{b_{12}\sqrt{\left( \frac{8b_{11}-81b_{22}}{b_{12}}%
				\right) ^{2}+64}} & \frac{8}{\sqrt{\left( \frac{8b_{11}-81b_{22}}{b_{12}}%
				\right) ^{2}+64}} & 0 \\
		0 & 1 & 0 \\
		0 & 0 & 1%
	\end{array}%
	\right)
\end{equation}%
For our calculations, the values of the free parameters $b_{ij}$ are
fixed by their values in the case of $\tan \beta =5$ as reported in
table (\ref{t8}). As a result, by inserting the expression of $\mathcal{Y}_{\nu }$ in eq. (%
\ref{cpa2}), we find that the $CP$ asymmetry parameter $\varepsilon
_{N_{3}^{c}}$ depends on the coupling constant $\lambda _{1}$, the
trimaximal mixing
parameters $\theta $ and $\sigma $, the Majorana $CP$ phases $\alpha _{31}$%
and $\alpha _{21}$ as well as the light neutrino masses $m_{i=1,2,3}$.
Approximately, $\varepsilon _{N_{3}^{c}}$ is given as%
\begin{eqnarray*}
	\varepsilon _{N_{3}^{c}} &=&10^{-6}\frac{\lambda _{1}^{2}}{8\pi }\left\{ %
	\left[ 6.\,\allowbreak 194\sin ^{4}\theta \sin ^{2}(\frac{\alpha _{31}}{2}%
	)\cos ^{2}\sigma \sin ^{2}\sigma \right] f\left( \frac{m_{1}}{m_{3}}\right)
	\right.  \\
	&&\left. +\left[ 2.\,\allowbreak 3226\sin ^{2}\theta \left( \cos \sigma \cos
	\left( \frac{\alpha _{21}-\alpha _{31}}{2}\right) +\sin \sigma \sin \left(
	\frac{\alpha _{21}-\alpha _{31}}{2}\right) \right) ^{2}\right] f\left( \frac{%
		m_{2}}{m_{3}}\right) \right\}
\end{eqnarray*}
Assuming that the coupling constant $\lambda _{1}$ is of order one and
taking into account the obtained regions of the parameters $\theta $\ and $%
\sigma $, the Majorana $CP$ phases $\alpha _{31}$ and $\alpha _{21}$ as well
as the neutrino masses $m_{i=1,2,3}$, the baryon asymmetry parameter $%
\left\vert \varepsilon _{N_{3}^{c}}\right\vert $ is up to order $\mathcal{O(}%
10^{-12}-10^{-10}\mathcal{)}$. However, as discussed in section
\ref{sec6}, to generate the observed baryon asymmetry, the parameter
$\left\vert
\varepsilon _{N_{3}^{c}}\right\vert $ must be of order $\mathcal{O(}10^{-5}%
\mathcal{)}$. Therefore, the charged lepton contribution to the $CP$
asymmetry parameter $\left\vert \varepsilon _{N_{3}^{c}}\right\vert
$\ is too small and subsequently the baryon asymmetry parameter
$Y_{B}$ is strongly suppressed.

Before we close this appendix, we discuss the possibility of
producing a successful leptogenesis using higher dimensional Yukawa
operators as an alternative to the additional coupling in eq.
(\ref{D}). Using the charge assignments of $D_{4}$ and $U(1)$
symmetries, we find that there are three invariant six dimensional
operators that can be used as a correction to the leading order
Dirac Yukawa matrix
\begin{equation}
	\frac{1}{\Lambda ^{2}}N_{3,2}^{c}F_{2,3}H_{5}\rho _{i}\xi _{2}\quad \text{%
		where}\mathrm{\quad }i=1,2,3  \label{six}
\end{equation}%
Since our model is renormalizable, these operators must be derived
from renormalizable Yukawa couplings involving the existing
messenger fields
listed in table (\ref{mes}). For example, generating the operator $\frac{1}{%
	\Lambda ^{2}}N_{3,2}^{c}F_{2,3}H_{5}\rho _{2}\xi _{2}$ calls for a
new messenger field $X_{6}$ which transforms as $SU(5)$ quintet,
$D_{4}$\ singlet $1_{++}$ and has a $U(1)$ charge equals to $-18$.
Nevertheless, the absence of this messenger field in our model
forbids the existence of this operator.\newline On the other hand,
even if we add messenger fields to allow the operators in eq.
(\ref{six}), we end up with a highly suppressed contribution to the
$CP$ asymmetry parameter $\left\vert \varepsilon
_{N_{3}^{c}}\right\vert $. As a verification, we use the same
example as above where the renormalizable
superpotential which induces the effective coupling $\frac{\lambda _{10}}{%
	\Lambda ^{2}}N_{3,2}^{c}F_{2,3}H_{5}\rho _{2}\xi _{2}$ is given by%
\begin{equation}
	W_{D}^{ren}=N_{3,2}^{c}F_{2,3}X_{5}+\overline{X}_{5}\rho _{2}X_{6}+\overline{%
		X}_{6}H_{5}\xi _{2},
\end{equation}%
where the coupling constants are omitted for simplicity. Subsequently, the total Yukawa mass matrix reads as%
\begin{equation}
	\mathcal{Y}_{D}=Y_{D}+\delta Y_{D}=\frac{m_{D}}{\upsilon
		_{u}}+\delta Y_{D}=\left(
	\begin{array}{ccc}
		\lambda _{1} & 0 & 0 \\
		0 & \lambda _{1} & 0 \\
		0 & 0 & \lambda _{1}%
	\end{array}%
	\right) +\frac{\left\vert \lambda _{10}\right\vert \upsilon _{\rho
			_{2}}\upsilon _{\xi _{2}}}{\Lambda ^{2}}\left(
	\begin{array}{ccc}
		0 & 0 & 0 \\
		0 & 0 & 1 \\
		0 & 1 & 0%
	\end{array}%
	\right) e^{i\phi _{H}}
\end{equation}%
where $\lambda _{10}$\ is a complex coupling constant $\lambda
_{10}=\left\vert \lambda _{10}\right\vert e^{i\phi _{H}}$. The $CP$
asymmetry parameter $\varepsilon _{N_{3}}$ corresponding to the
lightest RH neutrino $N_{3}$ is given approximately by
\begin{eqnarray}
	\varepsilon _{N_{3}} &\simeq &\frac{1}{9\pi }\left( \frac{\left\vert
		\lambda
		_{10}\right\vert \upsilon _{\rho _{2}}\upsilon _{\xi _{2}}}{\Lambda ^{2}}%
	\right) ^{2}\cos ^{2}\phi _{H}\left[ 2\sin ^{2}(2\theta )\sin ^{2}(\sigma -%
	\frac{\alpha _{31}}{2})f\left( \frac{m_{1}}{m_{3}}\right) \right.
	\label{eph} \\
	&&+\left. \sin ^{2}\theta \sin ^{2}\left( \sigma +\frac{(\alpha
		_{21}-\alpha _{31})}{2}\right) f\left( \frac{m_{2}}{m_{3}}\right)
	\right]   \nonumber
\end{eqnarray}%
The obtained $CP$ asymmetry parameter $\varepsilon _{N_{3}}$ is
proportional to the factor $\varepsilon _{N_{3}}\sim \left(
\frac{\left\vert \lambda
	_{10}\right\vert \upsilon _{\rho _{2}}\upsilon _{\xi _{2}}}{\Lambda ^{2}}%
\right) ^{2}$ which involves the flavon VEV $\frac{\upsilon _{\rho _{2}}}{%
	\Lambda }$ from the neutrino sector as well as $\frac{\upsilon _{\xi _{2}}}{%
	\Lambda }$ from the up-quark sector. According to the numerical
analysis we
have performed in the two sectors, we derive the interval of the ratio $%
\frac{\lambda _{7}\upsilon _{\rho _{2}}}{\Lambda }=-\frac{\mathrm{k}}{2}%
\simeq \lbrack -0.27557,0.26298]$ while the size of the flavon VEV\ $\frac{%
	\lambda _{12}^{u}\upsilon _{\xi _{2}}}{\Lambda }=a_{12}\simeq
0.1530\times
10^{-2}$ in the case of $\tan \beta =10$\footnote{%
	Notice that the value of the flavon VEV $\frac{\lambda
		_{12}^{u}\upsilon
		_{\xi _{2}}}{\Lambda }=a_{12}\simeq 0.15866\times 10^{-3}$ in the case of $%
	\tan \beta =5$ is much smaller and therefore the estimate on the $CP$ asymmetry parameter $\varepsilon _{N_{3}}$ becomes much
	suppressed.}. In order to get an estimate on the obtained $CP$
asymmetry parameter $\varepsilon _{N_{3}}$ in eq. (\ref{eph}), we
assume, as is reasonable to do, that the coupling constants $\lambda
_{7}$, $\lambda _{12}^{u}$ and $\left\vert \lambda _{10}\right\vert
$ are of order one and
we allow the phase $\phi _{H}$ to vary in the interval $[0,2\pi ]$%
. Therefore, we find that the $CP$ asymmetry parameter $\left\vert
\varepsilon _{N_{3}^{c}}\right\vert $ is up to order $\mathcal{O}%
(10^{-12}-10^{-7})$ which is too small to account for a successful
leptogenesis. As a result, the baryon asymmetry parameter $Y_{B}$ is
strongly suppressed when addressing leptogenesis through the six
dimensional operator $\frac{\lambda _{10}}{\Lambda
	^{2}}N_{3,2}^{c}F_{2,3}H_{5}\rho
_{2}\xi _{2}$. The same discussion holds for the other two operators $\frac{1%
}{\Lambda ^{2}}N_{3,2}^{c}F_{2,3}H_{5}\rho _{1}\xi _{2}$ and $\frac{1}{%
	\Lambda ^{2}}N_{3,2}^{c}F_{2,3}H_{5}\rho _{3}\xi _{2}$.
\section{Some aspects of the dihedral group $D_{4}$}
\label{app3}
The dihedral group $D_{4}$ is a finite group that is generated by the
reflection $t$ and the $45^{\circ }$ rotation $s$ satisfying $s^{4}=t^{2}=I$
and $tst=s^{-1}$. A rotation followed by a reflection is different than a
reflection followed by a rotation which means that the two generators $s$
and $t$ do not commute with each other. This non-Abelian group has $5$
irreducible representations $R_{i=1,...,5}$: one doublet denoted as $2_{0,0}$%
, and four singlets denoted as $1_{+,+}$(the trivial singlet), $1_{+,-}$ $%
1_{-,+}$ and $1_{-,-}$. The indices of these representations represent their
characters under the two generators $t$ and $s$ as in the following table
\begin{equation}
	\begin{tabular}{llllll}
		\hline\hline
		$\chi _{R_{i}}$ & $\chi _{2_{0,0}}$ & $\chi _{1_{+,+}}$ & $\chi _{1_{+,-}}$
		& $\chi _{1_{-,+}}$ & $\chi _{1_{-,-}}$ \\ \hline\hline
		$t$ & $0$ & $+1$ & $+1$ & $-1$ & $-1$ \\ \hline
		$s$ & $0$ & $+1$ & $-1$ & $+1$ & $-1$ \\ \hline\hline
	\end{tabular}%
\end{equation}%
The squares of the dimensions of these irreducible representations are
related to the order $8$ of the $D_{4}$ group through the formula $%
8=1_{+,+}^{2}+1_{+,-}^{2}+1_{-,+}^{2}+1_{-,-}^{2}+2_{0,0}$. Let us now turn
to the tensor products among the irreducible representations of $D_{4}$. The
tensor product between two $D_{4}$ doublets is decomposed into a sum of the
four singlet representations of $D_{4}$ as%
\begin{eqnarray}
	\left(
	\begin{array}{c}
		x_{1} \\
		x_{2}%
	\end{array}%
	\right) _{2_{0,0}}\otimes \left(
	\begin{array}{c}
		y_{1} \\
		y_{2}%
	\end{array}%
	\right) _{2_{0,0}} &=&\left( x_{1}y_{2}+x_{2}y_{1}\right) _{1_{+,+}}\oplus
	\left( x_{1}y_{1}+x_{2}y_{2}\right) _{1_{+,-}}\oplus \left(
	x_{1}y_{2}-x_{2}y_{1}\right) _{1_{-,+}}  \notag \\
	&&\oplus \left( x_{1}y_{1}-x_{2}y_{2}\right) _{1_{-,-}}  \label{c1}
\end{eqnarray}%
while the tensor products among the singlet representations can be expressed
as%
\begin{equation}
	1_{d,e}\otimes 1_{f,g}=1_{df,eg}\qquad \text{with}\qquad d,e,f,g=\pm
	\label{c2}
\end{equation}%
For more details on the $D_{4}$ group, see, e.g., \cite{A2}.
\section{Vacuum alignment of $D_{4}$ flavon doublets}
\label{app4}
Establishing an origin of the VEV directions is an essential part when using
non-Abelian discrete flavor symmetries to build models of fermion masses and
mixing. In our model, the VEVs of the $D_{4}$ doublet flavons pointing in
the directions given in eqs. (\ref{vq}) and (\ref{va}) were assumed in order
to produce the charged fermions and neutrino masses consistent with the
experimental data. One of the well-known approaches to check if these VEV
directions are a solution of the scalar potential is by
introducing a set of alignment fields called driving fields and a continuous
$U(1)_{R}$ symmetry. Under such a symmetry, the matter superfields
including right-handed neutrinos carry charge $+1$, flavons and
Higgs fields are uncharged while the driving fields have charge $+2$ \cite{C22}. As a result of the these $U(1)_{R}$ charge assignments, the driving fields couple only to flavons and appear linearly in
the superpotential, while the vacuum alignment is obtained by setting their
F-terms to zero. In general, the alignment through F-terms provide also
relations between flavons VEVs. Here, we introduce two driving fields
denoted as $\chi _{q}$ and $\chi _{\nu }$ transforming under $D_{4}\times
U(1)$ as%
\begin{equation}
	\chi _{q}\sim (\{1_{-,-}\},\{30\})\qquad ,\qquad \chi _{\nu }\sim
	(\{1_{-,+}\},\{-20\})
\end{equation}%
The renormalizable terms involving these driving fields invariant under the
flavor symmetry $D_{4}\times U(1)$ are given by%
\begin{equation}
	\mathcal{W}_{d}=c_{1}\chi _{q}\left( \Omega \Phi \right)
	_{1_{-,-}}+c_{2}\chi _{\nu }\left( \Gamma \digamma \right)
	_{1_{-,+}}+c_{3}\chi _{\nu }\left( \digamma \digamma \right)
	_{1_{-,+}}+c_{4}\chi _{\nu }\left( \Gamma \Gamma \right)
	_{1_{-,+}}+c_{5}\chi _{\nu }\rho _{2}\rho _{3}
\end{equation}%
In the SUSY limit where the F-terms of $\chi _{q}$ and $\chi _{\nu }$
vanish, the condition for the minima are%
\begin{eqnarray}
	\frac{\partial \mathcal{W}_{d}}{\partial \chi _{q}} &=&c_{1}\left( \Omega
	_{1}\Phi _{1}-\Omega _{2}\Phi _{2}\right) =0  \nonumber \\
	\frac{\partial \mathcal{W}_{d}}{\partial \chi _{\nu }} &=&c_{2}\left( \Gamma
	_{1}\digamma _{2}-\Gamma _{2}\digamma _{1}\right) +c_{3}\left( \digamma
	_{1}\digamma _{2}-\digamma _{2}\digamma _{1}\right) +c_{4}\left( \Gamma
	_{1}\Gamma _{2}-\Gamma _{2}\Gamma _{1}\right) +c_{5}\rho _{2}\rho
	_{3}=0  \label{df}
\end{eqnarray}%
Clearly, the first equation admits three different solutions given by%
\[
\begin{array}{ccccc}
	\left( 1\right)  & : & \left\langle \Phi \right\rangle =(\upsilon _{\Phi
	},\upsilon _{\Phi })^{T} & , & \left\langle \Omega \right\rangle =(\upsilon
	_{\Omega },\upsilon _{\Omega })^{T} \\
	\left( 2\right)  & : & \left\langle \Phi \right\rangle =(0,\upsilon _{\Phi
	})^{T} & , & \left\langle \Omega \right\rangle =(\upsilon _{\Omega },0)^{T}
	\\
	\left( 3\right)  & : & \left\langle \Phi \right\rangle =(\upsilon _{\Phi
	},0)^{T} & , & \left\langle \Omega \right\rangle =(0,\upsilon _{\Omega })^{T}%
\end{array}%
\]%
where the last solution is the one we have chosen to generate the Yukawa
matrices of the down-type quarks $\mathcal{Y}_{d}$ and charged leptons $%
\mathcal{Y}_{e}$ in eqs. (\ref{yd}) and (\ref{ye}), respectively. As for the
second equation in (\ref{df}), it admits the VEV direction given by
\begin{equation}
	\left\langle \rho _{2}\right\rangle =\upsilon _{\rho _{2}}\quad ,\quad
	\left\langle \rho _{3}\right\rangle =\upsilon _{\rho _{3}}\quad ,\quad
	\left\langle \digamma \right\rangle =(\upsilon _{\digamma },\upsilon
	_{\digamma })^{T}\quad ,\quad \left\langle \Gamma \right\rangle =(0,\upsilon
	_{\Gamma })^{T},
\end{equation}%
which we have used to produce the Majorana mass matrix provided the
following relation between the involved VEVs holds%
\begin{equation}
	\upsilon _{\digamma }=\frac{c_{5}}{c_{2}}\frac{\upsilon _{\rho _{2}}\upsilon
		_{\rho _{3}}}{\upsilon _{\Gamma }} \label{fe}
\end{equation}
According to the assumptions we have adopted to obtain the total neutrino mass matrix -- see eqs. (\ref{as1}) and (\ref{as2}) -- it follows that the set of flavon VEVs $\{\upsilon _{\rho _{1}}$,$%
\upsilon _{\digamma }\}$ and $\{\upsilon _{\rho _{2}},\upsilon
_{\rho _{3}},\upsilon _{\Gamma }\}$ are of the same order of
magnitude. Moreover, the flavon VEVs $\upsilon _{\rho _{2}}$ and $%
\upsilon _{\digamma }$ are related in eq. (\ref{fe}) through the couplings $%
c_{2}$ and $c_{5}$ which we assume that they are of the same order. As a result, we deduce that all the flavons used in
the neutrino sector are comparable to each other which is in
agreement with the numerical analysis performed in Section
\ref{sec5}.\newline On the contrary, the first equation in (\ref{df})
responsible for aligning the flavon doublets $\Omega $ and $\Phi $
does not induce any relation between their VEVs $\upsilon _{\Omega
}$ and $\upsilon _{\Phi }$. This is clearly reasonable since they
are not of the same order of magnitude as discussed numerically in
section \ref{sec5}. These two flavon VEVs contribute respectively to the
second and the third generations of down quarks (charged leptons)
which are strongly hierarchical.

						\end{document}